\documentclass[onecolumn]{aastex631}
\usepackage{amsmath,mathrsfs}

\received{2026 January 22}
\revised{2026 March 2}
\accepted{2026 March 2}
\published{2026 March 20}

\begin{document}

\title{The maximum mass and rotational kinetic energy of rapidly rotating neutron stars}

\author[0000-0001-9120-7733]{Shao-Peng Tang}
\affiliation{Key Laboratory of Dark Matter and Space Astronomy, Purple Mountain Observatory, Chinese Academy of Sciences, Nanjing 210033, China}
\author[0000-0002-7505-7795]{Yong-Jia Huang}
\affiliation{Key Laboratory of Dark Matter and Space Astronomy, Purple Mountain Observatory, Chinese Academy of Sciences, Nanjing 210033, China}
\affiliation{RIKEN Interdisciplinary Theoretical and Mathematical Sciences Program (iTHEMS), RIKEN, Wako 351-0198, Japan}
\author[0000-0002-8966-6911]{Yi-Zhong Fan}
\affiliation{Key Laboratory of Dark Matter and Space Astronomy, Purple Mountain Observatory, Chinese Academy of Sciences, Nanjing 210033, China}
\affiliation{School of Astronomy and Space Science, University of Science and Technology of China, Hefei, Anhui 230026, China}
\correspondingauthor{Yi-Zhong Fan}
\email{yzfan@pmo.ac.cn}

\begin{abstract}
Rapid uniformly-rotating neutron stars are expected to be formed for instance in the collapse of some massive stars, the accretion of compact object binaries, and double neutron star mergers. The huge amount of the rotational energy has been widely believed to be the source of some cosmic gamma-ray bursts and superluminous supernovae. Benefited from the constraints on the equation of state of the neutron star matter set by the latest multi-messenger data, the chiral effective field theory and perturbative quantum chromodynamics, here we present the maximum gravitational mass as well as the rotational energy for a neutron star at a given spin period. Our nonparametric equation of state analysis reveals that the critical Keplerian configurations ($\Omega_{\rm kep}^{\rm crit}=1.02_{-0.07}^{+0.06}\times 10^{4}~{\rm rad/s}$) can sustain maximum gravitational masses of $M_{\rm kep}^{\rm crit}=2.73 \pm 0.09 M_\odot$ with corresponding rotational energy reaching $E_{\rm rot,kep}^{\rm crit}=2.36^{+0.24}_{-0.22}\times 10^{53}$ erg. However, the maximum rotational energy that can be feasibly extracted from a neutron star is limited to $1.40^{+0.14}_{-0.13}\times 10^{53}$ erg, which holds for a baryon mass of $2.66^{+0.10}_{-0.09}M_\odot$. All these parameters, obtained via the nonparametric reconstruction of the equation of state, are at the $68.3\%$ confidence level and the adoption of a quarkonic model yields rather similar results. These findings are found to have already set some intriguing constraints on the millisecond magnetar interpretation of some exciting data.
\end{abstract}

\section{Introduction} \label{sec:intro}
In the universe, there are various kinds of very energetic outbursts such as gamma-ray bursts and superluminous supernovae. Though the black hole central engine has been widely adopted to interpret these phenomena, the rapidly rotating magnetized neutron star model can well account for some observational facts \citep{1992Natur.357..472U, 1992ApJ...392L...9D, 1998ApJ...505L.113K, 1998A&A...333L..87D, 2001ApJ...552L..35Z, 2006ChJAA...6..513G, 2010ApJ...717..245K,2010ApJ...719L.204W} and has also attracted wide attention.
One particularly interesting scenario is the merger of the double neutron stars, in which the formed massive remnant, if not collapsed into black hole quickly, will rotate at the mass shedding limit \citep{2011PhRvL.107e1102S, 2017RPPh...80i6901B}.
Very rapid rotation of the newly formed neutron star is also expected if the progenitor massive star has a high angular momentum \citep{2015Natur.523..189G}.
A fundamental question is hence how large the rotational kinetic energy of the extremely spinning neutron star could be.
An equally important issue is how much of this energy is extractable. We define the extractable rotational energy as the difference between a given spinning configuration and its terminal state along a constant baryonic mass sequence, motivated by the approximate conservation of baryonic mass during spin-down evolution. Depending on the stellar mass, the terminal state corresponds either to a non-rotating configuration or to the onset of collapse.
In principle, this energy reservoir governs the violence of the energetic explosions as well as the capability of accelerating the ultra-high energy cosmic rays.
The other relevant question is: what is the upper bound on the allowed gravitational mass in the case of rapid uniform rotation.
It plays an essential role in revealing the nature of the $\sim 2.4-3M_\odot$ mysterious compact objects detected in the Galaxy \citep{2024Sci...383..275B} and by the gravitational wave detectors \citep{2020ApJ...896L..44A, 2023PhRvX..13d1039A}.
Certainly, it is not possible to answer these two intriguing questions analytically.

The main purpose of this work is to reliably evaluate them numerically, benefiting from the recent breakthroughs on neutron star mass and radius measurements as well as the theoretical advancements.
For instance, the NICER mission has recently extended its measurements by determining the radius of PSR J0437–4715 \citep{2024ApJ...971L..20C} and PSR J0614–3329 \citep{2025ApJ...995...60M}, complementing earlier mass–radius determinations for PSR J0030+0451 \citep{2024ApJ...961...62V} and PSR J0740+6620 \citep{2024ApJ...974..294S, 2024ApJ...974..295D}.
These data, together with gravitational-wave event GW170817 \citep{2018PhRvL.121p1101A, 2019PhRvX...9a1001A}, we now have five neutron stars with reasonably measured mass and radius.
These multimessenger observations provide increasingly stringent constraints on the neutron-star equation of state \citep{2023ApJ...950..107G, 2023PhRvD.108i4014B, 2024PhRvD.109d3052F, 2025PhRvD.112h3009T}, and offer valuable insights into the composition of dense matter in the stellar core \citep{2020NatPh..16..907A, 2023SciBu..68..913H, 2023NatCo..14.8451A, 2026PhRvR...8a3253B}.
On the theoretical frontier, the chiral effective field theory ($\chi$EFT) can robustly describe the equation of state up to the nuclear saturation density \citep{2019PhRvL.122d2501D}, while perturbative quantum chromodynamic (pQCD) calculations have emerged as a valuable tool for constraining the very high-density behavior of nuclear matter \citep{2021PhRvL.127p2003G, 2022PhRvL.128t2701K, 2023ApJ...950..107G, 2023SciBu..68..913H}.
The measurements of the masses of about 100 neutron stars provide a mass distribution function that can be used to constrain $M_{\rm TOV}$, the maximum gravitational mass of nonrotating neutron stars \citep{2018MNRAS.478.1377A, 2020PhRvD.102f3006S, 2024PhRvD.109d3052F}.
Previously, the combination of these data and theoretical information has already yielded tight constraints on the equation of state of the dense neutron star matter and hence the accurate inference of $M_{\rm TOV}=2.25^{+0.08}_{-0.07}\,M_\odot$ \citep{2024PhRvD.109d3052F, 2024ApJ...974..244T} (see also, e.g., \citealt{2025PhRvD.112b3045B}).
Interestingly, such a maximum mass already seems to be challenged by the indirectly-measured gravitational mass $2.35\pm 0.17M_\odot$ of PSR J0952–0607 \citep{2022ApJ...934L..17R}.
This puzzle may be resolved, as the rapid rotation of PSR J0952–0607 could effectively enhance its gravitational mass \citep{1995ApJ...444..306S,2016MNRAS.459..646B}.

Motivated by the above facts, in this work we examine the effect of rapid uniform rotation, which is widely adopted in interpreting observational data in the literature.
Recent studies have investigated the rotational quasi-universal relations using nonparametric equation of state reconstructions or incorporating new theoretical inputs such as pQCD \citep{2024ApJ...962...61M, 2024PhRvD.109b3020L, 2025arXiv250911882K}.
In this work, we reconstruct the equation of state of the neutron star matter in two ways, one is the nonparametric Gaussian Process-based model and the other is a quarkyonic model.
The latter is physically motivated and has recently attracted considerable attention \citep{2023PhRvD.108e4013X, 2024PhRvC.109b5807P, 2025arXiv251023405K}. It also allows us to test the robustness of our results against different EOS modeling assumptions. Using EOSs consistent with current observational and theoretical constraints, we perform numerical calculations with the RNS code \citep{1995ApJ...444..306S}.
We derive the range of critical collapse masses as a function of rotation period and, concurrently, the corresponding range of maximum rotational energy and maximum extractable energy.
Our main finding includes the maximum angular velocity of $1.02_{-0.07}^{+0.06} \times 10^{4}~{\rm (rad/s)}$, the rotational kinetic energy bound of $2.36^{+0.24}_{-0.22} \times 10^{53}$ erg, the maximum extractable energy bound of $1.40^{+0.14}_{-0.13}\times 10^{53}$ erg, and the gravitational mass upper limit of $2.73 \pm 0.09 M_\odot$, for rapid uniformly-rotating neutron stars. The above values are obtained via the nonparametric reconstruction of the equation of state and the adoption of a quarkonic model yields rather similar results.
These thresholds are expected to be very helpful in better understanding some astrophysical phenomena.

\section{Methods}\label{sec:methods}
Previous studies employing nonparametric approaches—including the single-layer feed-forward neural network \citep{2021ApJ...919...11H, 2023SciBu..68..913H}, the piecewise linear sound speed model \citep{2022PhRvX..12a1058A, 2023ApJ...949...11J}, and Gaussian process (GP) method \citep{2019PhRvD..99h4049L, 2020PhRvD.101f3007E, 2020PhRvD.101l3007L}—have produced very consistent EOS reconstruction results \citep{2024PhRvD.109d3052F}. In this work, we adopt the GP method as our primary tool for constructing the neutron star EOS, which allows for a flexible, smooth extension from the low-density regime into the high-density domain while avoiding the limitations of fixed parametric forms (for further details, see e.g., \citealt{2023ApJ...950..107G, 2024PhRvD.109h3037T, 2025PhRvD.112f3003L, 2025PhRvD.112h3009T}). Specifically, we describe the sound speed via an auxiliary variable 
\begin{equation}
\phi(n) \equiv -\ln\!\left(1/c_s^2(n) - 1\right)\,,
\end{equation}
which is treated as a multivariate Gaussian distribution. That is, $\phi(n) \sim \mathcal{N}\big(-\ln(1/\bar{c}_s^2 - 1),\, K(n,n')\big)$, with a Gaussian kernel $K(n,n') = \eta \exp[-(n-n')^2/(2l^2)]$. The three hyperparameters of this GP, namely the variance $\eta$, correlation length $l$, and mean sound-speed-squared $\bar{c}_s^2$, are drawn from hyperprior distributions: $\eta \sim \mathcal{N}(1.25,\,0.2^2)$, $l \sim \mathcal{N}(0.5\,n_{\rm s},\,(0.25\,n_{\rm s})^2)$, and $\bar{c}_s^2 \sim \mathcal{N}(0.5,\,0.25^2)$, where $n_{\rm s}$ is nuclear saturation density. We condition the GP on the low-density $\chi$EFT EOS segment as follows. Let $\phi_{\rm CET}$ denote the values of $\phi(n)$ at densities $n_{\rm CET}\le 1.1\,n_{\rm s}$, drawn from the posterior samples of \citet{2019PhRvL.122d2501D}. We treat these as `data' points for the GP with an assumed variance $\sigma^2_{\phi_{\rm CET}} = 10^{-4}$ at each $n_{\rm CET}$. The conditioned GP then yields 
\begin{equation}
\phi_{\rm GP}^{*} \mid n_{\rm CET}, \phi_{\rm CET}, \sigma^2_{\phi_{\rm CET}}, n_{\rm GP} \sim \mathcal{N}\left(\bar{\phi}_{\rm GP}^{*}, {\rm cov}(\phi_{\rm GP}^{*})\right),
\end{equation}
where
\begin{equation}
\begin{aligned}
\bar{\phi}_{\rm GP}^* &= \bar{\phi}_{\rm GP} + K(n_{\rm GP}, n_{\rm CET})\big[K(n_{\rm CET}, n_{\rm CET}) \\
&+ \sigma^2_{\phi_{\rm CET}} I\big]^{-1}(\phi_{\rm CET} - \bar{\phi}_{\rm GP})\,, \\
{\rm cov}(\phi_{\rm GP}^{*}) &= K(n_{\rm GP}, n_{\rm GP}) - K(n_{\rm GP}, n_{\rm CET})\big[ \\
&K(n_{\rm CET}, n_{\rm CET}) + \sigma^2_{\phi_{\rm CET}} I\big]^{-1}K(n_{\rm CET}, n_{\rm GP})\,.
\end{aligned}
\end{equation}
In the above, $\bar{\phi}_{\rm GP} = -\ln(1/\bar{c}_s^2 - 1)$, $I$ is the identity matrix, and $n_{\rm GP}$ are logarithmically spaced between $1.1\,n_{\rm s}$ and $12\,n_{\rm s}$. Each EOS sample is generated by first drawing a set of GP hyperparameters from the hyperpriors, then sampling a realization of $\phi_{\rm GP}^*(n)$ from the conditioned GP, and finally solving for the corresponding sound speed $c_s(n)$ and integrating to obtain the pressure $p(n)$ and energy density $\varepsilon(n)$.

In addition to this flexible, phenomenological GP model, we implement a physically motivated parameterized EOS model, the quarkyonic model \citep{2019PhRvL.122l2701M, 2024PhRvL.132k2701F}, which offers microscopic insight into high-density matter and naturally explains the stiffening of the EOS. In the quarkyonic model, as the baryon density exceeds a threshold $n_t$, nucleons become confined to a narrow shell in momentum space near the Fermi surface while free quarks populate the lower momentum states. This coexistence of a free quark Fermi sea with nucleons contributes extra kinetic pressure, resulting in a rapid increase in pressure and a peak in the speed of sound. In the formulation of \citet{2020PhRvD.102b3021Z}, an explicit momentum cutoff is imposed on nucleons. In practice, the minimum momentum for neutrons (or protons) is parameterized as
\begin{equation}
k_{0(n,p)} = k_{F(n,p)}\left[1 - \left(\frac{\Lambda}{k_{F(n,p)}}\right)^\alpha - \frac{\kappa_{n,p}\Lambda}{9\,k_{F(n,p)}}\right],
\end{equation}
where $k_{F(n,p)}$ is the Fermi momentum of neutrons (or protons), $\Lambda$ controls the width of the momentum shell, $\kappa_{n,p}$ are model parameters determined by the ambient beta-equilibrium conditions at $n_t$, and $\alpha$ (typically 2 or 3) determines the sharpness of the cutoff. This construction forces quarks to ``drip'' out of the nucleons, thereby increasing their kinetic energy and overall pressure. Moreover, the model is extended to include protons and leptons so that chemical and beta equilibrium are maintained. While the complete formulation involves several parameters to accurately reproduce nuclear matter properties, we here adopt a simplified ndu version from \citet{2020PhRvD.102b3021Z}. This reduced version retains the key physical ingredients—the nucleon–quark transition, momentum-space restrictions, and equilibrium conditions—while allowing for efficient, analytic EOS calculations. The prior distributions for the ndu model parameters are chosen based on physical expectations and experimental constraints. Specifically, $L$ (the slope of symmetry energy), $\Lambda$, and $n_t$ are uniformly distributed within ranges of $(20, 100){\rm MeV}$, $(10, 2500){\rm MeV}$, and $(0.18, 0.8){\rm fm}^{-3}$, respectively. The binding energy ${\rm BE}$ and symmetry energy parameter $S_v$ are Gaussian distributed, with ${\rm BE} \sim \mathcal{N}(-15.97, 0.2^2){\rm MeV}$ and $S_v \sim \mathcal{N}(32, 0.55^2){\rm MeV}$, respectively \citep{2024PhRvC.110d4320D}. The PNM potential parameter $\gamma_1$ in Equation~(10) of \citet{2020PhRvD.102b3021Z} is uniformly distributed, $\gamma_1 \sim \text{Uniform}(0.5, 5)$. And we employ the SLy4 \citep{2001A&A...380..151D} crust EOS for $n<0.5n_{\rm s}$. In contrast to the GP construction, we do not impose $\chi$EFT constraints on the quarkyonic model in the present analysis, as the two frameworks are implemented independently. We employ \textsc{Bilby} \citep{2019ApJS..241...27A} with \textsc{Dynesty} \citep{2020MNRAS.493.3132S} for nested sampling, adopting an evidence tolerance of $\Delta\log Z=0.1$ and 2000 live points (other sampler settings follow the \textsc{Bilby} defaults).

Once the EOS is constructed by either method, we solve the Tolman–Oppenheimer–Volkoff (TOV) and Regge–Wheeler equations to generate mass–radius and mass–tidal deformability curves.

We use the latest multimessenger observations and theoretical pQCD calculations to rigorously constrain the neutron star equation of state. The key datasets incorporated into our Bayesian framework are as follows: First, the NICER X-ray timing observations provide precise radius measurements for several neutron stars, including PSR J0437–4715 with a radius of $11.36^{+0.95}_{-0.63}$ km (at the $68\%$ credible level) \citep{2024ApJ...971L..20C}, PSR J0030+0451 with a radius of $11.71^{+0.88}_{-0.83}$ km derived using the ST+PDT model \citep{2024ApJ...961...62V} (with alternative models like PDT-U being disfavored \citep{2024ApJ...966...98L}), PSR J0740+6620 with a radius of $12.49^{+1.28}_{-0.88}$ km \citep{2024ApJ...974..294S}, and most recently PSR J0614–3329 with $R = 10.29^{+1.01}_{-0.86}$ km \citep{2025ApJ...995...60M}. Second, the gravitational-wave constraints from GW170817 provide critical information on the tidal deformability of neutron stars \citep{2018PhRvL.121p1101A,2019PhRvX...9a1001A}. Third, we adopt the maximum mass determinations based on the marginalized posterior distribution of the maximum mass cutoff for neutron stars as reported in \citet{2024PhRvD.109d3052F}. Finally, we apply high-density pQCD constraints at $10n_{\rm s}$ \citep{2023ApJ...950..107G}, which, though not direct observations, provide important theoretical boundary conditions for our EOS framework.

After constraining the equation of state using observational data and theoretical constraints via both the quarkyonic and GP-based methods, the resulting EOSs are used as inputs for the RNS code \citep{1995ApJ...444..306S} to compute the rotational properties of neutron stars. Specifically, we calculate sequences of uniformly rotating neutron stars at various spins up to the mass-shedding limit for each EOS. Within these sequences, we identify the turning point along a constant-angular-momentum sequence, which marks the onset of secular instability and defines the critical configuration. At this configuration, the star is expected to collapse, and various quantities, including gravitational mass ($M_{\rm crit}$), baryonic mass, radius, angular momentum, moment of inertia ($I$), and angular velocity ($\Omega$), are outputted by the RNS code. We compute the rotational energy as $E_{\rm rot} = \frac{1}{2} I \Omega^2$, and also evaluate the extractable rotational energy, $E_{\rm ext}$, along constant-rest-mass sequences. For baryonic masses below the TOV limit ($M_{\rm b,TOV}$), $E_{\rm ext}$ equals the total rotational energy at a given angular velocity. For baryonic masses exceeding $M_{\rm b,TOV}$ (i.e., the SMNS scenario), $E_{\rm ext}$ is defined as the difference between the rotational energy at a given spin and that at the point of collapse.

\section{Results} \label{sec:results}
\begin{figure}
    \centering
    \includegraphics[width=0.5\textwidth]{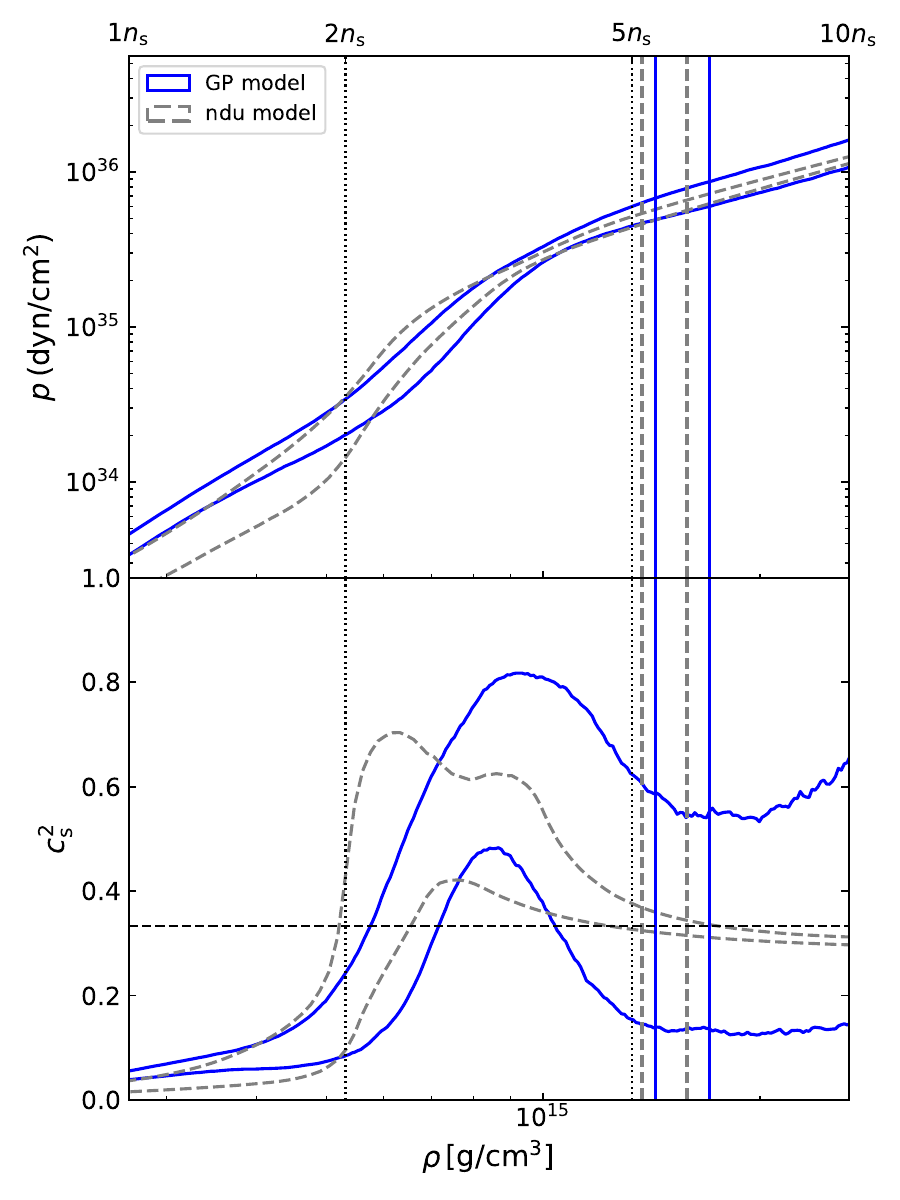}
    \caption{The equations of state (the upper panel) and the squared sound speeds (the lower panel) reconstructed with the GP model (blue) and a specific quarkyonic model (gray) \citep{2020PhRvD.102b3021Z}.}
    \label{fig:EoS}
\end{figure}
The reconstructed equations of state are reported in Figure~\ref{fig:EoS} and throughout this work the regions are at $68.3\%$ confidence level. We find that the quarkyonic model exhibits an earlier peak in the sound speed compared to the GP framework. At densities above $\sim 3n_{\rm s}$, the resulting $\rho-p$ relations are broadly consistent with those obtained from the GP method, although the quarkyonic model yields narrower uncertainty bands. The reduced uncertainties, with the upper boundary approaching the conformal limit at the core densities of massive neutron stars, generally make the quarkyonic EOS stiffer in the intermediate-density regime. Consequently, the inferred radius of a $1.4\,M_\odot$ NS, $R_{1.4}=11.78^{+0.42}_{-0.37}$ km, is consistent with that obtained from the GP model ($R_{1.4} = 11.87^{+0.47}_{-0.45}$ km). In contrast, the radius of a $2\,M_\odot$ NS, $R_{2.0}=12.05^{+0.42}_{-0.41}$ km, is likely larger than that predicted by the GP model ($R_{2.0} = 11.83^{+0.51}_{-0.43}$ km).

\begin{figure}
    \centering
    \includegraphics[width=0.5\textwidth]{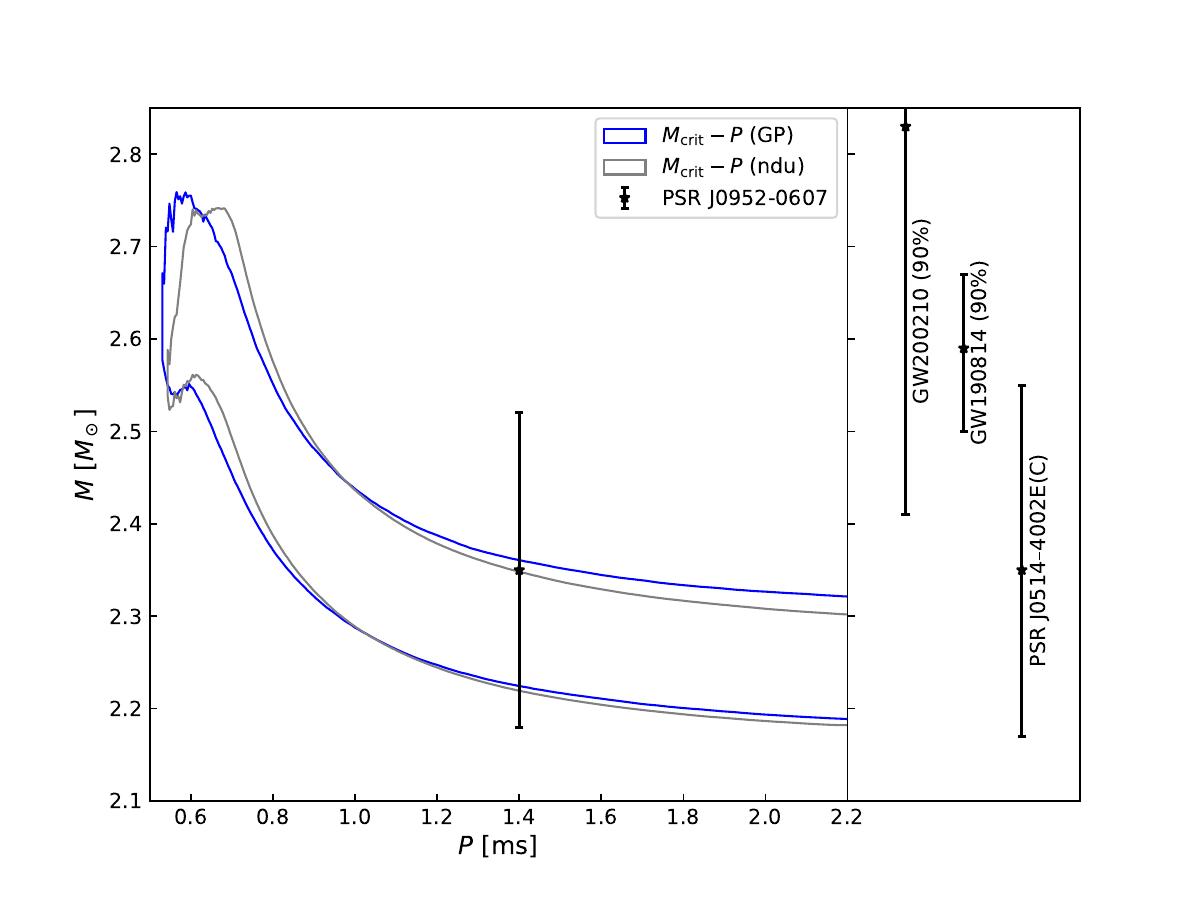}
    \caption{Critical collapse mass $M_{\rm crit}$, as a function of rotation period $P$. The blue and gray curves correspond to the GP and ndu models, respectively. The data of PSR J0952–0607 \citep{2022ApJ...934L..17R}, the companion of PSR J0514–4002E \citep{2024Sci...383..275B} and the secondary stars of GW190814 and GW200210 \citep{2023PhRvX..13d1039A} are shown for comparison.}
    \label{fig:mcrit-p}
\end{figure}
Despite differences in the reconstructed equations of state from the two models, their predicted rotational limits are remarkably similar, as demonstrated by the allowed range of the critical mass $M_{\rm crit}$ as a function of spin period shown in Figure~\ref{fig:mcrit-p}.
When $P\gg 1$ ms, the allowed maximum mass approaches $M_{\rm TOV}$, while for very rapid rotation ($P<1$ ms) the maximum mass can be significantly enhanced.
It turns out that the massive PSR J0952–0607 \citep{2022ApJ...934L..17R}, if confirmed by the future direct measurement, is consistent with the current knowledge learned from the multi-messenger data.
The companion of PSR J0514–4002E has a best-fit gravitational mass of $2.35^{+0.20}_{-0.18}M_\odot$ \citep{2024Sci...383..275B}, which can be either a neutron star or a low mass black hole because of the lack of other information.
The secondary star of GW190814 is however too massive to be a neutron star unless the spin is as quick as $\leq 1$ ms (see also e.g., \citealt{2020MNRAS.499L..82M, 2021Univ....7..182L}).
Even assuming a zero gravitational wave radiation, the strength of the dipole magnetic field should be $\leq 10^{9}$ Gauss to keep such a quick rotation in a typical merger timescale of $\sim 1$ Gyr.
However, it is widely anticipated that the neutron star born with $P\leq 1$ ms would be accompanied by very strong dipole magnetic field \citep{1992ApJ...392L...9D, 2006Sci...312..719P}.
We thus suggest that the secondary star of GW190814 is most likely a low mass black hole. 

\begin{figure}
    \centering
    \includegraphics[width=0.5\textwidth]{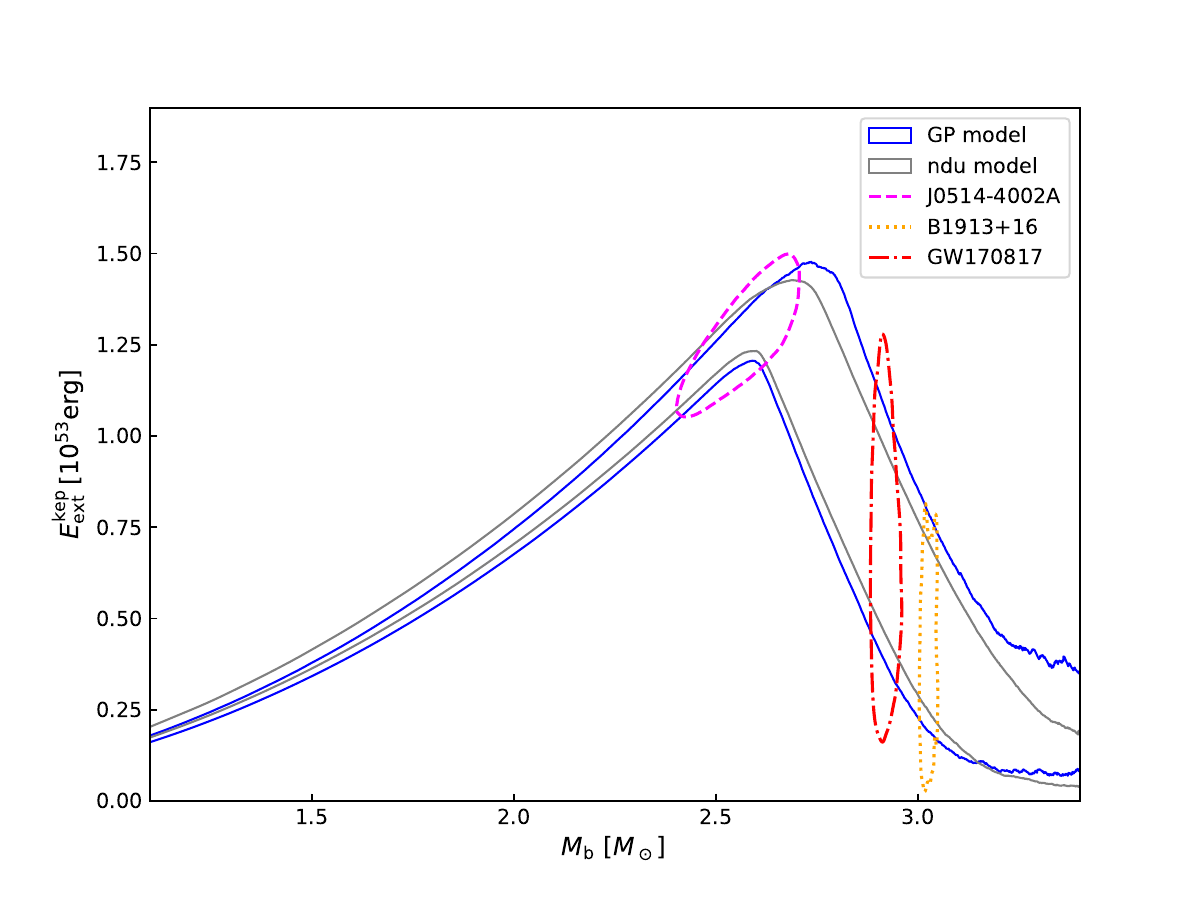}
    \caption{Extractable rotational energy of NS at mass-shedding limit, $M_{\rm ext}^{\rm kep}$, as a function of baryonic mass. The blue and gray curves correspond to the GP and ndu models, respectively. Expected values for J0514–4002A \citep{2019MNRAS.490.3860R}, B1913+16 \citep{2010ApJ...722.1030W}, and GW170817 \citep{2019PhRvX...9c1040A} are presented.}
    \label{fig:eext-mb}
\end{figure}

For some supramassive neutron stars, the rotational energy can not be efficiently extracted due to collapse. 
For this reason, we introduce the term {\it extractable} rotational energy to reflect the capability of powering the violent explosion by a living neutron star. Figure~\ref{fig:eext-mb} shows the variation of the extractable rotational energy from NSs rotating at the Keplerian limit, $E_{\rm ext}^{\rm kep}$, as a function of baryonic mass. We find that $E_{\rm ext}^{\rm kep}$ reaches its maximum at the TOV-limit baryonic mass $M_{\rm b,TOV}$. The corresponding values are $M_{\rm b,TOV} = 2.66^{+0.10}_{-0.09}~M_\odot$, $E_{\rm ext}^{\rm kep} = 1.40^{+0.14}_{-0.13} \times 10^{53}~\mathrm{erg}$ for the GP model, and $M_{\rm b,TOV} = 2.64^{+0.08}_{-0.07}~M_\odot$, $E_{\rm ext}^{\rm kep} = 1.40^{+0.08}_{-0.10} \times 10^{53}~\mathrm{erg}$ for the quarkyonic (ndu) model. For baryonic masses below $M_{\rm b,TOV}$, $E_{\rm ext}^{\rm kep}$ increases with mass. However, for supramassive NSs with $M_{\rm b} > M_{\rm b,TOV}$, collapse to BHs occurs once sufficient rotational energy is lost, resulting in a decline of $E_{\rm ext}^{\rm kep}$ with increasing mass. We also compute the remnant baryonic masses for two Galactic BNS systems (assuming they will merge in the future), as well as for the GW170817 event, and predict their corresponding maximum extractable rotational energies. In these estimates, we assume $0.1\,M_\odot$ of material remains outside the remnant. J0514-4002A is expected to have $E_{\rm ext}^{\rm kep}\approx 1.25\times10^{53}$ erg, indicating that the merger remnants of the light double NS systems can indeed serve as one of the sources of the most violent phenomena. However, in heavy systems the extractable energy is significantly reduced due to earlier collapse to black holes. Interestingly, this even happens to GW170817, indicating a quick collapse, in agreement with the argument based on the lack of helium in the spectrum of AT2017gfo \citep{2024arXiv241103427S} (see however \citealt{2018ApJ...861..114Y}).

\begin{figure}
    \centering
    \includegraphics[width=0.5\textwidth]{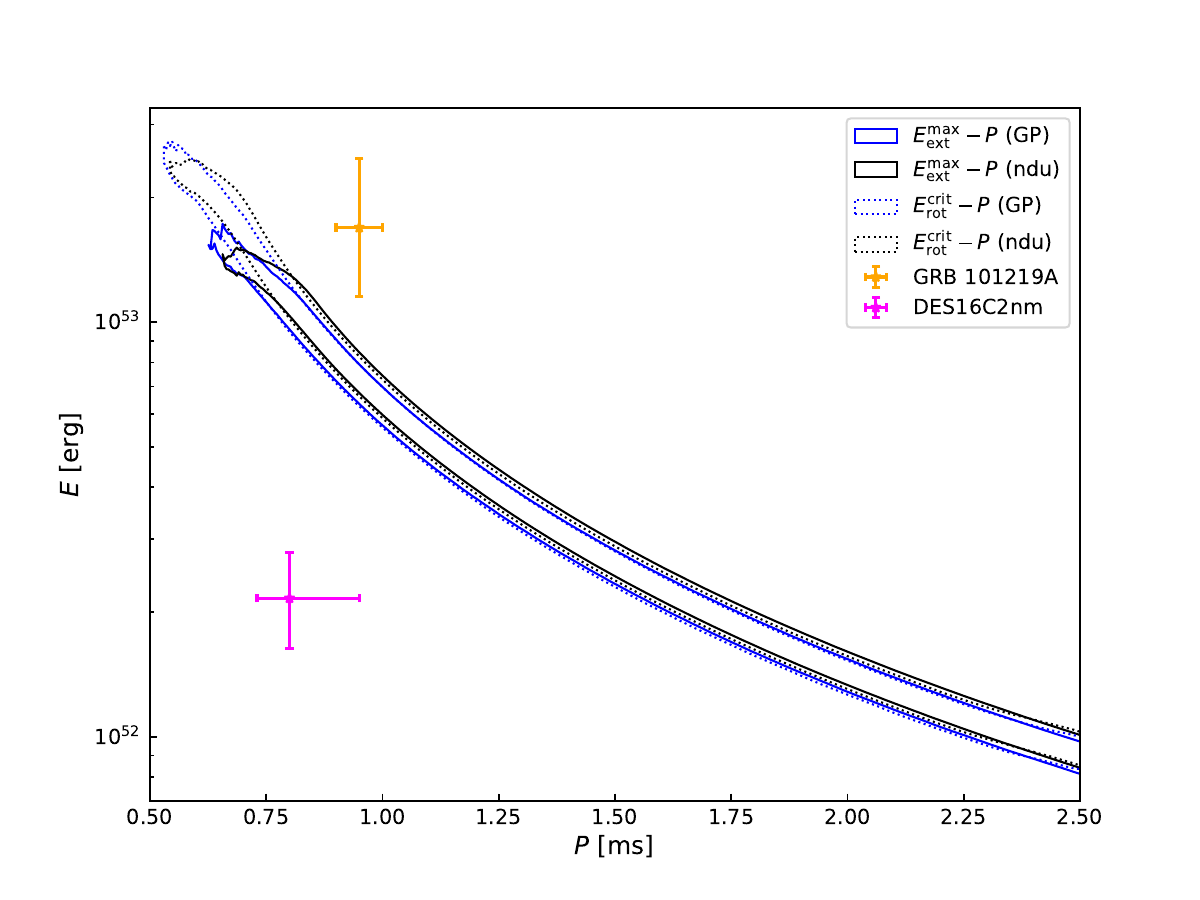}
    \caption{The solid lines show the maximum extractable rotational energy, $E_{\rm ext}^{\rm max}$, for neutron stars with baryonic mass equal to $M_{\rm b,TOV}$, as a function of rotation period $P$. The dotted lines represent the maximum rotational energy of neutron stars at a given period. Blue and black curves correspond to the GP and ndu models, respectively. The inferred parameters of DES16C2nm \citep{2021ApJ...921..180H} and GRB 101219A \citep{2013MNRAS.430.1061R} are shown for comparison.}
    \label{fig:eext-p}
\end{figure}

Since the rest-mass sequence at $M_{\rm b,TOV}$ yields the largest extractable rotational energy, we further examine its dependence on spin. In Figure~\ref{fig:eext-p}, the solid lines present the predicted ranges of the maximum extractable energy, $E_{\rm ext}^{\rm max}$ (for $M_{\rm b} = M_{\rm b,TOV}$), as a function of spin period $P$, based on our EOS ensemble. For comparison, we also present the critical rotational energy of the neutron stars (i.e., the dotted lines). The solid lines overlap with the dotted lines except at the highest spins, for which the neutron stars are supramassive.
A specific superluminoius supernova, DES16C2nm, at a redshift of 1.998, is highlighted because of the rather short period $P=0.80^{+0.15}_{-0.07}$ ms as well as the very high rotational kinetic energy of $\approx 3\times 10^{52}$ erg \citep{2021ApJ...921..180H}.
This value is already comparable with, though still below, our expected $E_{\rm ext}^{\rm max}$ at such a short period, suggesting that DES16C2nm could be among the most energetic superluminous supernovae in recent decades. 
CDF-S XT2, a $\sim 10^{3}$ s long X-ray transient with a peak luminosity of $\sim 3\times 10^{45}~{\rm erg~s^{-1}}$, has been attributed to the radiation of the non-collapsing magnetar formed in a double neutron star merger \citep{2019Natur.568..198X}.
Evidently, unless the X-ray emission efficiency is as low as $\sim 10^{-5}$ or most of the energy has been radiated in gravitational wave \citep{2013PhRvD..88f7304F, 2017ApJ...835..181L}, the magnetar interpretation has a serious missing energy problem.
Another direct application of our Figure~\ref{fig:eext-p} is on the magnetar interpretation of the X-ray plateau of GRB 101219A, which requires a rotation period of $0.95$ ms and the energy of $1.7\times 10^{53}$ erg \citep{2013MNRAS.430.1061R}, significantly exceeding our predicted $E_{\rm ext}^{\rm max}(P=0.95~{\rm ms})$.
This suggests that a magnetar central engine interpretation for GRB 101219A may not be tenable, and alternative mechanisms such as energy extraction from a newborn black hole via the Blandford-Znajek process \citep{2018MNRAS.475..266Z}, refreshed shocks \citep{1998ApJ...496L...1R}, structured or multi-component jets \citep{2004ApJ...605..300H}, or a coasting external shock phase \citep{2012ApJ...744...36S} could potentially explain this event.

\section{Summary and Discussion} \label{sec:summary}
In this work, we explored the maximum mass, rotational energy and extractable energy of rapidly rotating neutron stars, using a range of equations of state constrained by the latest multimessenger data.
We find that the EOS constraints inferred from the GP and quarkyonic frameworks are broadly consistent, although the quarkyonic model typically exhibits an earlier peak in the sound speed.
We computed sequences of uniformly rotating neutron stars, extending to the mass-shedding limit, and examined how the critical collapse mass, total rotational energy, and extractable energy vary with spin period.
Despite differences in the sound-speed behavior between the two EOS frameworks, the resulting rotational properties are remarkably similar. This consistency indicates that our conclusions are robust against the choice of EOS modeling strategy.
We then compared the constrained ranges of the critical collapse mass with observed data, such as the measured masses of PSR J0952$-$0607 and the companion of PSR J0514$-$4002E.
It turns out that both objects may be supra-massive neutron stars, stabilized by rapid rotation.
But for the secondary star of GW190814, it is too massive and a neutron star is only possible with an almost mass-shedding rotation as well as an extremely low dipole magnetic field.
As for the energy budget, though many superluminous supernovae have been detected and the magnetar model has been adopted to fit the data, the inferred highest rotational energy is $\sim 3\times 10^{52}$ erg for a $P\sim 0.8$ ms \citep{2021ApJ...921..180H}, which lies well below our derived bound and likely indicates a gravitational mass considerably smaller than $M_{\rm kep}^{\rm crit}$.
However, for GRB 101219A, the claimed $P\sim 0.95$ ms with a rotational energy of $\sim 1.7\times 10^{53}$ erg \citep{2013MNRAS.430.1061R} is in tension with our $E_{\rm ext}^{\rm max}$ limit (see Figure \ref{fig:eext-p}).
We also caution that the interpretation of short GRB data using the merger-formed magnetar model often suffers from a significant energy deficit problem, unless most of the angular momentum, and thus the rotational energy, of the nascent remnant is efficiently carried away by gravitational waves \citep{2013PhRvD..88f7304F}.
We note that our analysis is restricted to uniformly rotating, cold neutron stars, and does not explicitly account for the effects of differential rotation or finite temperature. This approximation is well justified for long-lived magnetars, where differential rotation is expected to be rapidly suppressed by strong magnetic braking and viscous processes, and thermal effects become negligible on cooling timescales. However, in the immediate aftermath of a binary neutron star merger, the remnant is likely to be both differentially rotating and thermally supported, which may temporarily enhance the maximum supported mass \citep{2014ApJ...790...19K, 2021PhRvC.103e5811K, 2021ApJ...912...69K, 2026PhRvD.113b3013T}. Such short-lived post-merger effects, while potentially important on dynamical timescales, are beyond the scope of the present work and will be explored in future studies.
Our final remark is that the remnants formed in the mergers of the lightest neutron star binaries are most likely supramassive or even stable, and the energy reservoir represented by $E_{\rm ext}^{\rm max}$ will be sufficient to produce extremely luminous counterparts and accelerate a large amount of ultra-high energy cosmic rays. The detection of such violent events in the future would further strength our conclusion.

\begin{acknowledgments}
This work is supported by the National Natural Science Foundation of China under Grants No. 12233011 and No. 12303056, the Project for Young Scientists in Basic Research (No. YSBR-088) of the Chinese Academy of Sciences, and the Postdoctoral Fellowship Program of China Postdoctoral Science Foundation (GZC20241915).
\end{acknowledgments}

\appendix

In this Appendix, we provide additional details regarding the posterior distributions of the quarkyonic model parameters, a comparison of the bulk properties of neutron stars (NSs) derived from the Gaussian process (GP) and quarkyonic models, and the reconstructed mass-radius relations. These analyses complement the main text and offer further insights into the equations of state (EOSs) constrained by recent multi-messenger NS observations and theoretical advancements. Using these constrained EOSs, we subsequently determine the rotational properties of neutron stars. Finally, we explore universal relations for the critical rotational energy derived from posterior samples of the GP model.

\begin{figure}
    \centering
    \includegraphics[width=0.48\textwidth]{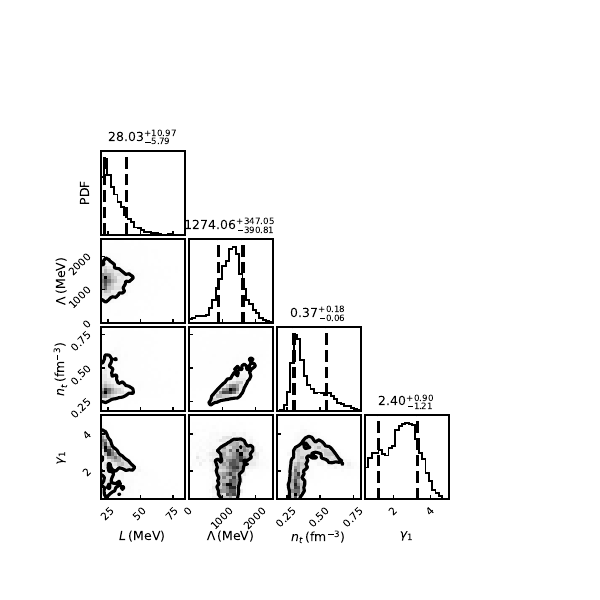}
    \caption{Posterior distributions of key parameters from the quarkyonic model, including the slope of symmetry energy $L$, the parameter $\Lambda$, the threshold density $n_t$, and the potential parameter $\gamma_1$. The values and their $68.3\%$ credible intervals are shown for each parameter.}
    \label{fig:eos-param}
\end{figure}
Figure \ref{fig:eos-param} displays the posterior distributions of key parameters from the quarkyonic model, including the slope of the symmetry energy $L$, the parameter $\Lambda$, the threshold density $n_t$, and the potential parameter $\gamma_1$. Notably, the slope of the symmetry energy $L$ is concentrated at the lower edge, with a value of $28.03_{-5.79}^{+10.97}$ MeV, consistent with constraints from nuclear experiments \citep{2017ApJ...848..105T, 2020PhRvL.125t2702D}. Additionally, the threshold density $n_t$ is well-constrained to $0.37^{+0.18}_{-0.06}~{\rm fm}^{-3}$, indicating a rapid stiffening that occurs at approximately twice the nuclear saturation density.
\begin{figure}
    \centering
    \includegraphics[width=0.48\textwidth]{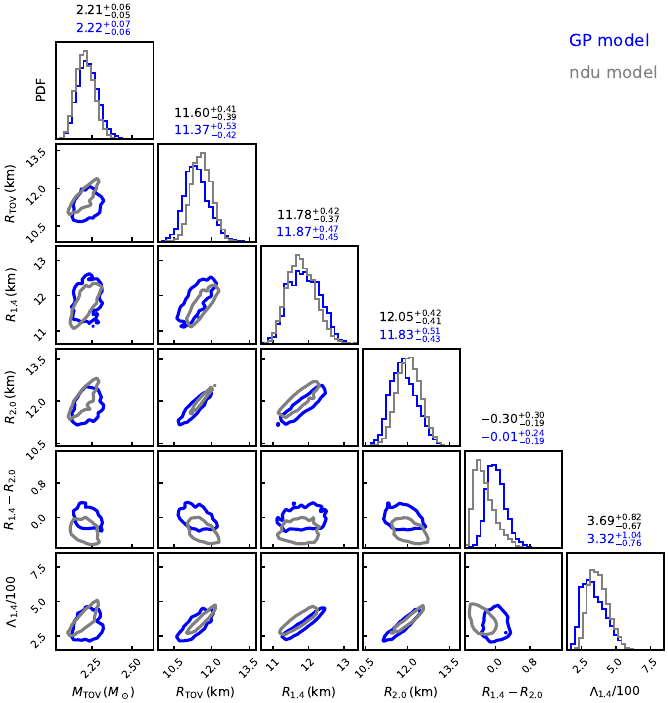}
    \caption{Comparison of the bulk properties of NSs derived from the GP (blue) and quarkyonic (gray) models. The corresponding $68.3\%$ credible intervals are indicated above the diagonal subplots.}
    \label{fig:bulk-prop}
\end{figure}
Figure \ref{fig:bulk-prop} compares the bulk properties of NSs derived from the GP and quarkyonic models, including the maximum mass ($M_{\rm TOV}$) and corresponding radius ($R_{\rm TOV}$) of a nonrotating NS, the radius ($R_{1.4}$) and tidal deformability ($\Lambda_{1.4}$) of a canonical $1.4\,M_\odot$ NS, the radius of a $1.4\,M_\odot$ NS ($R_{2.0}$), and the difference between $R_{1.4}$ and $R_{2.0}$. Both models yield consistent results, particularly for $M_{\rm TOV}$, with the quarkyonic model tending to favor slightly larger radii.
\begin{figure}
    \centering
    \includegraphics[width=0.48\textwidth]{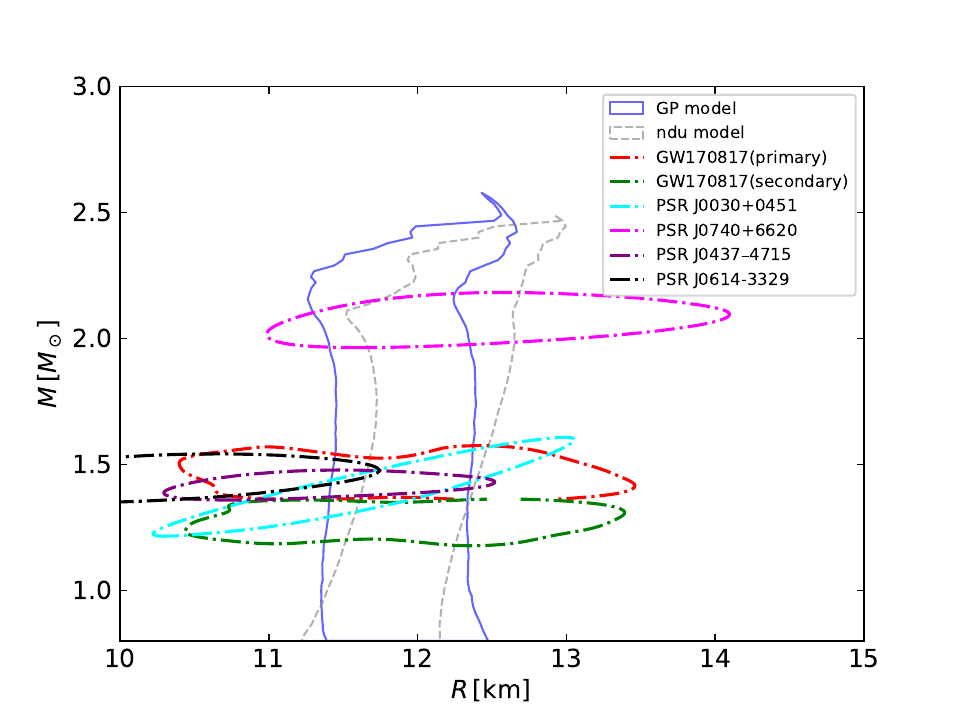}
    \caption{The $68.3\%$ credible mass-radius intervals reconstructed using the GP (blue) and quarkyonic (gray) models. The mass-radius data for PSR J0030+0451, PSR J0740+6620, PSR J0437–4715, PSR J0614–3329, and the components of the GW170817 event are shown for comparison.}
    \label{fig:mr-bound}
\end{figure}
Figure \ref{fig:mr-bound} shows the mass-radius relations for neutron stars reconstructed using the GP and quarkyonic models, along with the corresponding observational data for various neutron stars and gravitational wave events. The contours represent the $68.3\%$ credible regions for both models, with blue denoting the GP model and gray denoting the quarkyonic model. The mass-radius relations predicted by the two models are broadly similar, with the quarkyonic model suggesting slightly larger radii across the entire mass range. However, this is still much smaller than the results found in \citet{2024PhRvC.109b5807P} (with $R_{1.4} \sim 13.44$ km). This discrepancy may arise because the nondynamical quarkyonic model refined by \citet{2020PhRvD.102b3021Z} includes beta equilibrium, and the adoption of the ndu model, though simplified, provides a more accurate representation, leading to better alignment with the observed NS radii.

\begin{figure}
    \centering
    \includegraphics[width=0.48\textwidth]{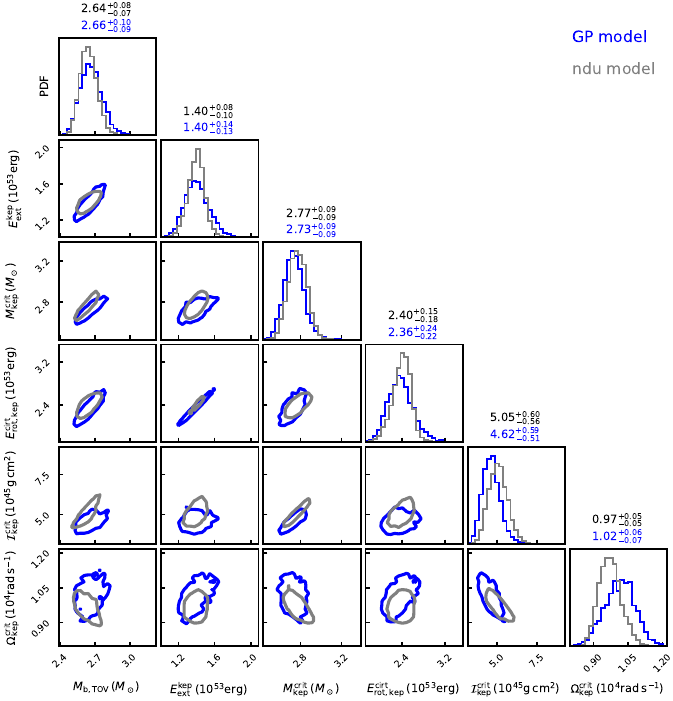}
    \caption{Corner plot showing joint posterior distributions of the maximum baryonic mass of nonrotating neutron stars $M_{\rm b,TOV}$, the corresponding extractable energy at the Keplerian limit $E_{\rm ext}^{\rm kep}$, the critical gravitational mass at the Keplerian limit $M_{\rm kep}^{\rm crit}$, the corresponding rotational energy $E_{\rm rot,kep}^{\rm crit}$, moment of inertia $\mathcal{I}_{\rm kep}^{\rm crit}$, and angular velocity $\Omega_{\rm kep}^{\rm crit}$. Values displayed above the diagonal panels indicate the $68\%$ credible intervals for each parameter.}
    \label{fig:rot-prop}
\end{figure}
Figure~\ref{fig:rot-prop} summarizes the key rotational properties of neutron stars. For a star with baryonic mass equal to the nonrotating maximum ($M_{\rm b,TOV}$), the extractable energy at the Keplerian limit $E_{\rm ext}^{\rm kep}$ is estimated to be $1.40^{+0.14}_{-0.13}\times10^{53}$ erg (GP) and $1.40^{+0.08}_{-0.10}\times10^{53}$ erg (ndu). The critical gravitational mass at the Keplerian limit $M_{\rm kep}^{\rm crit}$ is similar between models, with $2.73\pm0.09\,M_\odot$ (GP) and $2.77\pm0.09\,M_\odot$ (ndu). The ratio $M_{\rm kep}^{\rm crit}/M_{\rm TOV}$ is found to be $1.23\pm0.02$ for the GP model and $1.25\pm0.02$ for the ndu model, consistent with previous studies indicating a quasi-universal value close to $1.2$ \citep{2016MNRAS.459..646B, 2020PhRvD.101f3029S, 2024ApJ...962...61M}. The critical rotational energy $E_{\rm rot,kep}^{\rm crit}$ is $2.36^{+0.24}_{-0.22}\times10^{53}$ erg (GP) and $2.40^{+0.15}_{-0.18}\times10^{53}$ erg (ndu). The moment of inertia $\mathcal{I}_{\rm kep}^{\rm crit}$ differs slightly, with GP predicting $4.62^{+0.59}_{-0.51}\times10^{45}$ g cm$^2$ and ndu $5.05^{+0.60}_{-0.56}\times10^{45}$ g cm$^2$. Finally, the angular velocity $\Omega_{\rm kep}^{\rm crit}$ at the Keplerian limit is higher for the GP model, $1.02^{+0.06}_{-0.07}\times10^4$ rad s$^{-1}$, compared to $0.97^{+0.05}_{-0.05}\times10^4$ rad s$^{-1}$ for the ndu model. The results reported above exhibit relatively small uncertainties. For instance, the uncertainty in $M_{\rm kep}^{\rm crit}$ is comparable to that of $M_{\rm TOV}$ reported in \citet{2024PhRvD.109d3052F}, which is expected given that current neutron-star observations and theoretical inputs have already placed strong constraints on the EOSs.

\begin{figure}
    \centering
    \includegraphics[width=0.48\textwidth]{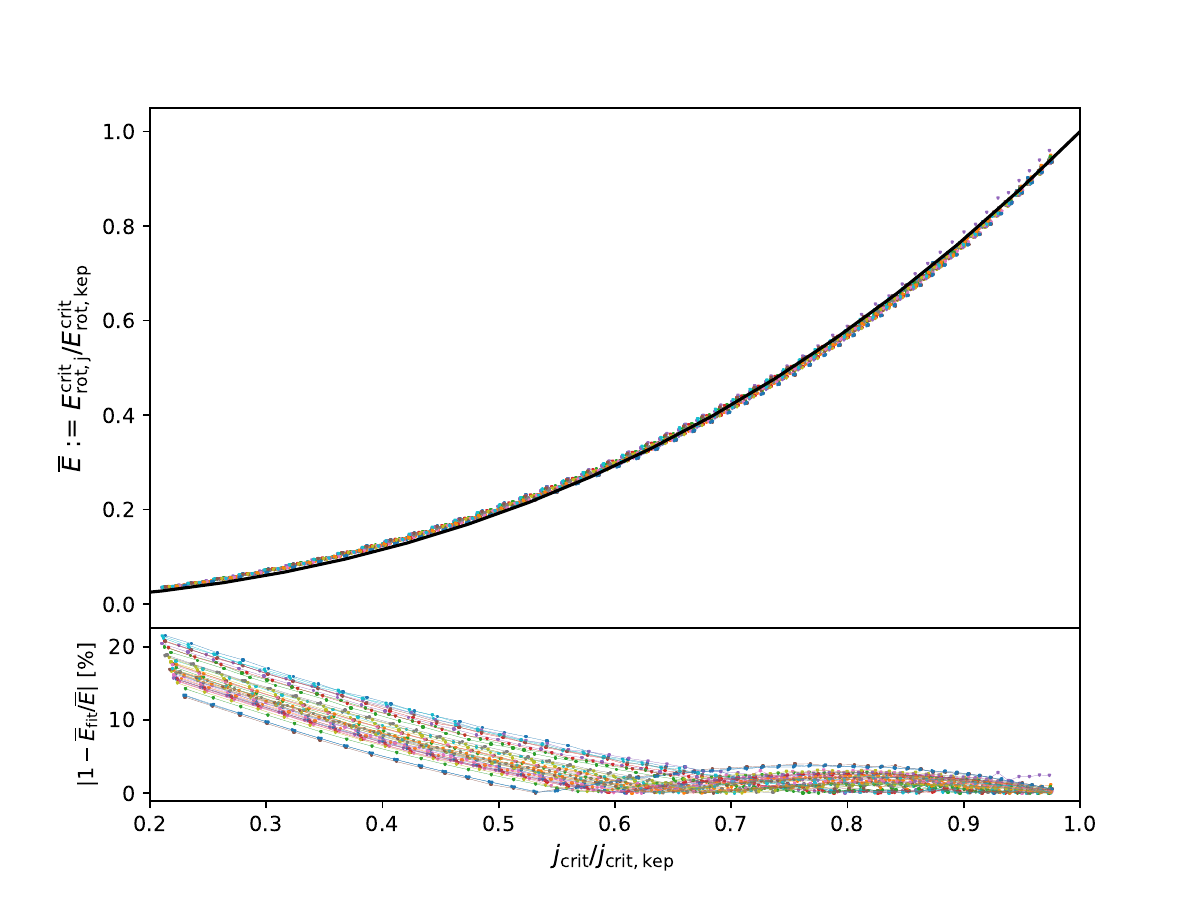}
    \caption{Normalized critical rotational energy, $\overline{E} = E_{\rm rot,j}^{\rm crit}/E_{\rm rot,kep}^{\rm crit}$, plotted as a function of the normalized spin parameter, $j_{\rm crit}/j_{\rm crit,kep}$. Colored points in the top panel represent the rotational properties of neutron stars computed from posterior samples of our EOS ensemble, while the solid black curve shows a polynomial fit to the data. The bottom panel displays the fractional residual, $\bigl|1 - \overline{E}_{\rm fit}/\overline{E}\bigr|$.}
    \label{fig:erot-jkep}
\end{figure}
We also update some universal relations regarding the total rotational energy, $E_{\rm rot}$. As shown in Figure~\ref{fig:erot-jkep}, the normalized critical rotational energy, $\overline{E} = E_{\rm rot,j}^{\rm crit}/E_{\rm rot,kep}^{\rm crit}$ (where $E_{\rm rot,kep}^{\rm crit}$ is the critical rotational energy at the Kepler limit), exhibits an excellent correlation with the normalized dimensionless angular momentum, $j_{\rm crit}/j_{\rm crit,kep}$. The solid black curve represents a polynomial fit, $y = 0.53\,x^2 + 0.47\,x^3$, and the lower panel displays the fractional residual. The scatter of points reflects the uncertainty across our EOS ensemble, illustrating that despite these uncertainties, the normalized critical rotational energy follows a smooth, predictable trend as the star’s spin approaches the mass-shedding limit.
\begin{figure}
    \centering
    \includegraphics[width=0.48\textwidth]{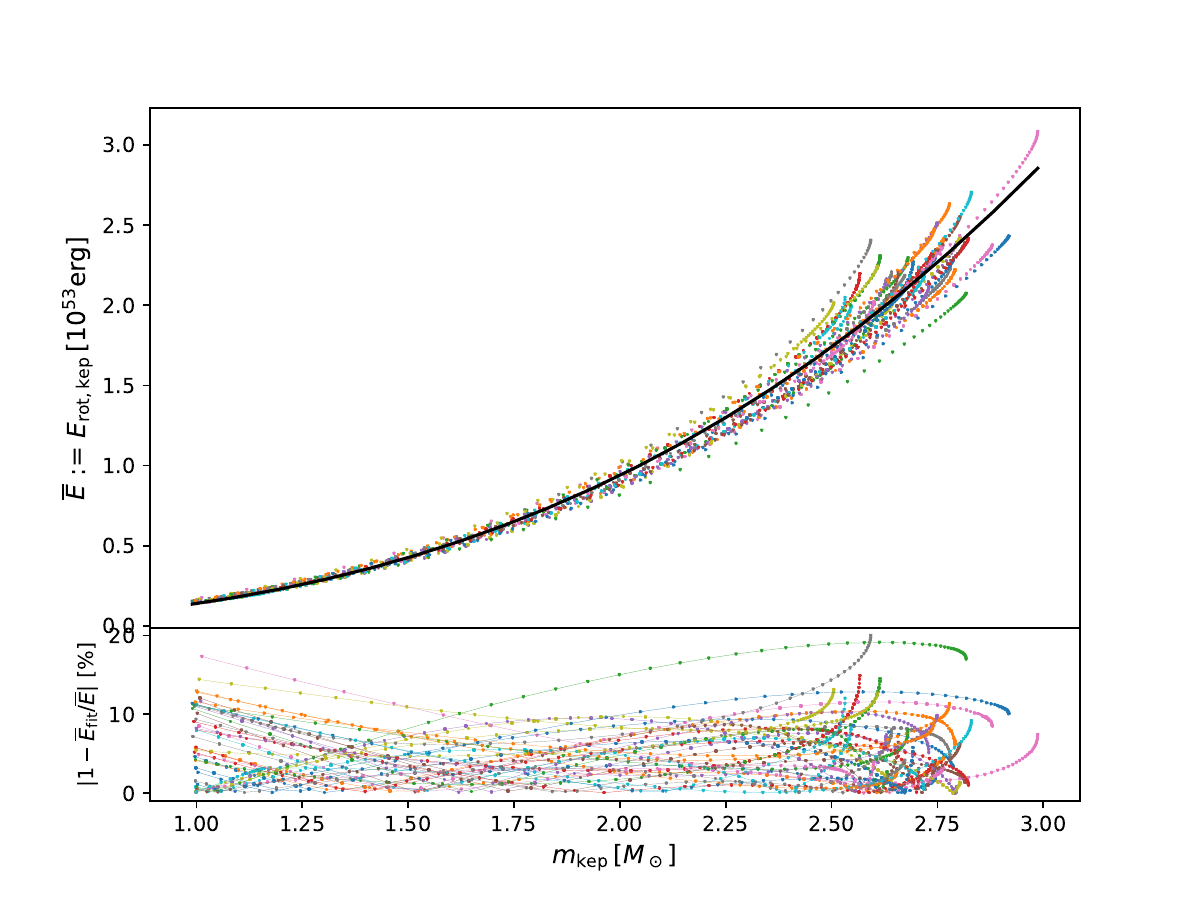}
    \caption{Rotational energy at the Kepler limit, $E_{\rm rot,kep}$ (in $10^{53}\,\mathrm{erg}$), as a function of the neutron star mass $m_{\rm kep}$. The black line is a polynomial fit and the bottom panel displays the fractional residual.}
    \label{fig:erot-mkep}
\end{figure}
We further examine the correlation between the rotational energy at the mass-shedding limit, $E_{\rm rot,kep}$, and the corresponding stellar mass, $m_{\rm kep}$. Figure~\ref{fig:erot-mkep} displays this relationship: the horizontal axis represents $m_{\rm kep}$ (in $M_\odot$), while the vertical axis shows $E_{\rm rot,kep}$ in units of $10^{53} \mathrm{erg}$. A polynomial fit, $y = 0.08\,x^3 + 0.08\,x^2 - 0.03\,x$, captures the trend well, with the bottom panel depicting the relative residual. This systematic behavior indicates that although different EOSs produce variations in the maximum mass at the mass-shedding limit, the underlying functional dependence between $E_{\rm rot,kep}$ and $m_{\rm kep}$ remains robust. These universal relations are valuable for estimating the rotational energy reservoir of a supramassive neutron star formed after a binary neutron star merger, even without precise knowledge of neutron star EOS.

\bibliography{refs.bib}{}

@ARTICLE{2014ApJ...790...19K,
       author = {{Kaplan}, J.~D. and {Ott}, C.~D. and {O'Connor}, E.~P. and {Kiuchi}, K. and {Roberts}, L. and {Duez}, M.},
        title = "{The Influence of Thermal Pressure on Equilibrium Models of Hypermassive Neutron Star Merger Remnants}",
      journal = {\apj},
         year = 2014,
        month = jul,
       volume = {790},
       number = {1},
          eid = {19},
        pages = {19},
          doi = {10.1088/0004-637X/790/1/19},
archivePrefix = {arXiv},
       eprint = {1306.4034},
 primaryClass = {astro-ph.HE},
       adsurl = {https://ui.adsabs.harvard.edu/abs/2014ApJ...790...19K}
}

@ARTICLE{2022PhRvX..12a1058A,
       author = {{Annala}, Eemeli and {Gorda}, Tyler and {Katerini}, Evangelia and {Kurkela}, Aleksi and {N{\"a}ttil{\"a}}, Joonas and {Paschalidis}, Vasileios and {Vuorinen}, Aleksi},
        title = "{Multimessenger Constraints for Ultradense Matter}",
      journal = {Physical Review X},
         year = 2022,
        month = jan,
       volume = {12},
       number = {1},
          eid = {011058},
        pages = {011058},
          doi = {10.1103/PhysRevX.12.011058},
archivePrefix = {arXiv},
       eprint = {2105.05132},
 primaryClass = {astro-ph.HE},
       adsurl = {https://ui.adsabs.harvard.edu/abs/2022PhRvX..12a1058A}
}

@ARTICLE{2025PhRvD.112b3045B,
       author = {{Biswas}, Bhaskar and {Rosswog}, Stephan},
        title = "{Simultaneously constraining the neutron star equation of state and mass distribution through multimessenger observations and nuclear benchmarks}",
      journal = {\prd},
         year = 2025,
        month = jul,
       volume = {112},
       number = {2},
          eid = {023045},
        pages = {023045},
          doi = {10.1103/8lv3-1ywb},
archivePrefix = {arXiv},
       eprint = {2408.15192},
 primaryClass = {astro-ph.HE},
       adsurl = {https://ui.adsabs.harvard.edu/abs/2025PhRvD.112b3045B}
}

@ARTICLE{2026PhRvD.113b3013T,
       author = {{Tsiopelas}, Stefanos and {Sedrakian}, Armen and {Oertel}, Micaela},
        title = "{Rapidly rotating hot nuclear and hypernuclear compact stars: Integral parameters and universal relations}",
      journal = {\prd},
         year = 2026,
        month = jan,
       volume = {113},
       number = {2},
          eid = {023013},
        pages = {023013},
          doi = {10.1103/zsxf-87zc},
archivePrefix = {arXiv},
       eprint = {2510.16239},
 primaryClass = {astro-ph.HE},
       adsurl = {https://ui.adsabs.harvard.edu/abs/2026PhRvD.113b3013T}
}

@ARTICLE{2020PhRvD.101f3029S,
       author = {{Shao}, Dong-Sheng and {Tang}, Shao-Peng and {Sheng}, Xin and {Jiang}, Jin-Liang and {Wang}, Yuan-Zhu and {Jin}, Zhi-Ping and {Fan}, Yi-Zhong and {Wei}, Da-Ming},
        title = "{Estimating the maximum gravitational mass of nonrotating neutron stars from the GW170817/GRB 170817A/AT2017gfo observation}",
      journal = {\prd},
         year = 2020,
        month = mar,
       volume = {101},
       number = {6},
          eid = {063029},
        pages = {063029},
          doi = {10.1103/PhysRevD.101.063029},
archivePrefix = {arXiv},
       eprint = {1912.08122},
 primaryClass = {astro-ph.HE},
       adsurl = {https://ui.adsabs.harvard.edu/abs/2020PhRvD.101f3029S}
}

@ARTICLE{2018MNRAS.475..266Z,
       author = {{Zhang}, Q. and {Lei}, W.~H. and {Zhang}, B.~B. and {Chen}, W. and {Xiong}, S.~L. and {Song}, L.~M.},
        title = "{Search for the signatures of a new-born black hole from the collapse of a supra-massive millisecond magnetar in short GRB light curves}",
      journal = {\mnras},
         year = 2018,
        month = mar,
       volume = {475},
       number = {1},
        pages = {266-276},
          doi = {10.1093/mnras/stx3229},
archivePrefix = {arXiv},
       eprint = {1712.04103},
 primaryClass = {astro-ph.HE},
       adsurl = {https://ui.adsabs.harvard.edu/abs/2018MNRAS.475..266Z}
}

@ARTICLE{1998ApJ...496L...1R,
       author = {{Rees}, M.~J. and {M{\'e}sz{\'a}ros}, P.},
        title = "{Refreshed Shocks and Afterglow Longevity in Gamma-Ray Bursts}",
      journal = {\apjl},
         year = 1998,
        month = mar,
       volume = {496},
       number = {1},
        pages = {L1-L4},
          doi = {10.1086/311244},
archivePrefix = {arXiv},
       eprint = {astro-ph/9712252},
 primaryClass = {astro-ph},
       adsurl = {https://ui.adsabs.harvard.edu/abs/1998ApJ...496L...1R}
}

@ARTICLE{2004ApJ...605..300H,
       author = {{Huang}, Y.~F. and {Wu}, X.~F. and {Dai}, Z.~G. and {Ma}, H.~T. and {Lu}, T.},
        title = "{Rebrightening of XRF 030723: Further Evidence for a Two-Component Jet in a Gamma-Ray Burst}",
      journal = {\apj},
         year = 2004,
        month = apr,
       volume = {605},
       number = {1},
        pages = {300-306},
          doi = {10.1086/382202},
archivePrefix = {arXiv},
       eprint = {astro-ph/0309360},
 primaryClass = {astro-ph},
       adsurl = {https://ui.adsabs.harvard.edu/abs/2004ApJ...605..300H}
}

@ARTICLE{2012ApJ...744...36S,
       author = {{Shen}, Rongfeng and {Matzner}, Christopher D.},
        title = "{Coasting External Shock in Wind Medium: An Origin for the X-Ray Plateau Decay Component in Swift Gamma-Ray Burst Afterglows}",
      journal = {\apj},
         year = 2012,
        month = jan,
       volume = {744},
       number = {1},
          eid = {36},
        pages = {36},
          doi = {10.1088/0004-637X/744/1/36},
archivePrefix = {arXiv},
       eprint = {1109.3453},
 primaryClass = {astro-ph.HE},
       adsurl = {https://ui.adsabs.harvard.edu/abs/2012ApJ...744...36S}
}

@ARTICLE{2025arXiv251023405K,
       author = {{Kalita}, Probit J and {Malik}, Tuhin and {Zhao}, Tianqi and {Kumar}, Bharat and {Lattimer}, James M.},
        title = "{Observable Signatures of a Quarkyonic Phase in Neutron Stars}",
      journal = {arXiv e-prints},
         year = 2025,
        month = oct,
          eid = {arXiv:2510.23405},
        pages = {arXiv:2510.23405},
          doi = {10.48550/arXiv.2510.23405},
archivePrefix = {arXiv},
       eprint = {2510.23405},
 primaryClass = {nucl-th},
       adsurl = {https://ui.adsabs.harvard.edu/abs/2025arXiv251023405K}
}

@ARTICLE{2023PhRvD.108e4013X,
       author = {{Xia}, Cheng-Jun and {Jin}, Hao-Miao and {Sun}, Ting-Ting},
        title = "{Quarkyonic matter and quarkyonic stars in an extended relativistic mean field model}",
      journal = {\prd},
         year = 2023,
        month = sep,
       volume = {108},
       number = {5},
          eid = {054013},
        pages = {054013},
          doi = {10.1103/PhysRevD.108.054013},
archivePrefix = {arXiv},
       eprint = {2307.03032},
 primaryClass = {hep-ph},
       adsurl = {https://ui.adsabs.harvard.edu/abs/2023PhRvD.108e4013X}
}

@ARTICLE{2023NatCo..14.8451A,
       author = {{Annala}, Eemeli and {Gorda}, Tyler and {Hirvonen}, Joonas and {Komoltsev}, Oleg and {Kurkela}, Aleksi and {N{\"a}ttil{\"a}}, Joonas and {Vuorinen}, Aleksi},
        title = "{Strongly interacting matter exhibits deconfined behavior in massive neutron stars}",
      journal = {Nature Communications},
         year = 2023,
        month = dec,
       volume = {14},
          eid = {8451},
        pages = {8451},
          doi = {10.1038/s41467-023-44051-y},
archivePrefix = {arXiv},
       eprint = {2303.11356},
 primaryClass = {astro-ph.HE},
       adsurl = {https://ui.adsabs.harvard.edu/abs/2023NatCo..14.8451A}
}

@ARTICLE{2020NatPh..16..907A,
       author = {{Annala}, Eemeli and {Gorda}, Tyler and {Kurkela}, Aleksi and {N{\"a}ttil{\"a}}, Joonas and {Vuorinen}, Aleksi},
        title = "{Evidence for quark-matter cores in massive neutron stars}",
      journal = {Nature Physics},
         year = 2020,
        month = jun,
       volume = {16},
       number = {9},
        pages = {907-910},
          doi = {10.1038/s41567-020-0914-9},
archivePrefix = {arXiv},
       eprint = {1903.09121},
 primaryClass = {astro-ph.HE},
       adsurl = {https://ui.adsabs.harvard.edu/abs/2020NatPh..16..907A}
}

@ARTICLE{2026PhRvR...8a3253B,
       author = {{Bauswein}, Andreas and {Nikolaidis}, Aristeidis and {Lioutas}, Georgios and {Kochankovski}, Hristijan and {Char}, Prasanta and {Mondal}, Chiranjib and {Oertel}, Micaela and {Tolos}, Laura and {Chamel}, Nicolas and {Goriely}, Stephane},
        title = "{Stellar properties indicating the presence of hyperons in neutron stars}",
      journal = {Physical Review Research},
         year = 2026,
        month = mar,
       volume = {8},
       number = {1},
          eid = {013253},
        pages = {013253},
          doi = {10.1103/ygtr-ktqk},
archivePrefix = {arXiv},
       eprint = {2507.10372},
 primaryClass = {astro-ph.HE},
       adsurl = {https://ui.adsabs.harvard.edu/abs/2026PhRvR...8a3253B}
}

@ARTICLE{2023PhRvD.108i4014B,
       author = {{Brandes}, Len and {Weise}, Wolfram and {Kaiser}, Norbert},
        title = "{Evidence against a strong first-order phase transition in neutron star cores: Impact of new data}",
      journal = {\prd},
         year = 2023,
        month = nov,
       volume = {108},
       number = {9},
          eid = {094014},
        pages = {094014},
          doi = {10.1103/PhysRevD.108.094014},
archivePrefix = {arXiv},
       eprint = {2306.06218},
 primaryClass = {nucl-th},
       adsurl = {https://ui.adsabs.harvard.edu/abs/2023PhRvD.108i4014B}
}

@ARTICLE{2024ApJ...962...61M,
       author = {{Musolino}, Carlo and {Ecker}, Christian and {Rezzolla}, Luciano},
        title = "{On the Maximum Mass and Oblateness of Rotating Neutron Stars with Generic Equations of State}",
      journal = {\apj},
         year = 2024,
        month = feb,
       volume = {962},
       number = {1},
          eid = {61},
        pages = {61},
          doi = {10.3847/1538-4357/ad1758},
archivePrefix = {arXiv},
       eprint = {2307.03225},
 primaryClass = {gr-qc},
       adsurl = {https://ui.adsabs.harvard.edu/abs/2024ApJ...962...61M}
}

@ARTICLE{2024PhRvD.109b3020L,
       author = {{Legred}, Isaac and {Sy-Garcia}, Bubakar O. and {Chatziioannou}, Katerina and {Essick}, Reed},
        title = "{Assessing equation of state-independent relations for neutron stars with nonparametric models}",
      journal = {\prd},
         year = 2024,
        month = jan,
       volume = {109},
       number = {2},
          eid = {023020},
        pages = {023020},
          doi = {10.1103/PhysRevD.109.023020},
archivePrefix = {arXiv},
       eprint = {2310.10854},
 primaryClass = {astro-ph.HE},
       adsurl = {https://ui.adsabs.harvard.edu/abs/2024PhRvD.109b3020L}
}

@ARTICLE{2025arXiv250911882K,
       author = {{Kr{\"u}ger}, Christian J. and {Celato}, Mariachiara},
        title = "{Universal relations for fast rotating neutron stars without equation of state bias}",
      journal = {arXiv e-prints},
         year = 2025,
        month = sep,
          eid = {arXiv:2509.11882},
        pages = {arXiv:2509.11882},
          doi = {10.48550/arXiv.2509.11882},
archivePrefix = {arXiv},
       eprint = {2509.11882},
 primaryClass = {gr-qc},
       adsurl = {https://ui.adsabs.harvard.edu/abs/2025arXiv250911882K}
}

@ARTICLE{2021ApJ...919...11H,
       author = {{Han}, Ming-Zhe and {Jiang}, Jin-Liang and {Tang}, Shao-Peng and {Fan}, Yi-Zhong},
        title = "{Bayesian Nonparametric Inference of the Neutron Star Equation of State via a Neural Network}",
      journal = {\apj},
         year = 2021,
        month = sep,
       volume = {919},
       number = {1},
          eid = {11},
        pages = {11},
          doi = {10.3847/1538-4357/ac11f8},
archivePrefix = {arXiv},
       eprint = {2103.05408},
 primaryClass = {hep-ph},
       adsurl = {https://ui.adsabs.harvard.edu/abs/2021ApJ...919...11H}
}

@ARTICLE{2021PhRvC.103e5811K,
       author = {{Khadkikar}, Sanika and {Raduta}, Adriana R. and {Oertel}, Micaela and {Sedrakian}, Armen},
        title = "{Maximum mass of compact stars from gravitational wave events with finite-temperature equations of state}",
      journal = {\prc},
         year = 2021,
        month = may,
       volume = {103},
       number = {5},
          eid = {055811},
        pages = {055811},
          doi = {10.1103/PhysRevC.103.055811},
archivePrefix = {arXiv},
       eprint = {2102.00988},
 primaryClass = {astro-ph.HE},
       adsurl = {https://ui.adsabs.harvard.edu/abs/2021PhRvC.103e5811K}
}

@ARTICLE{2021ApJ...912...69K,
       author = {{Koliogiannis}, P.~S. and {Moustakidis}, Ch. C.},
        title = "{Thermodynamical Description of Hot, Rapidly Rotating Neutron Stars, Protoneutron Stars, and Neutron Star Merger Remnants}",
      journal = {\apj},
         year = 2021,
        month = may,
       volume = {912},
       number = {1},
          eid = {69},
        pages = {69},
          doi = {10.3847/1538-4357/abe542},
archivePrefix = {arXiv},
       eprint = {2007.10424},
 primaryClass = {astro-ph.HE},
       adsurl = {https://ui.adsabs.harvard.edu/abs/2021ApJ...912...69K}
}

@ARTICLE{2019PhRvL.122d2501D,
       author = {{Drischler}, C. and {Hebeler}, K. and {Schwenk}, A.},
        title = "{Chiral Interactions up to Next-to-Next-to-Next-to-Leading Order and Nuclear Saturation}",
      journal = {\prl},
         year = 2019,
        month = feb,
       volume = {122},
       number = {4},
          eid = {042501},
        pages = {042501},
          doi = {10.1103/PhysRevLett.122.042501},
       adsurl = {https://ui.adsabs.harvard.edu/abs/2019PhRvL.122d2501D}
}

@ARTICLE{2024PhRvC.110d4320D,
       author = {{Drischler}, C. and {Giuliani}, P.~G. and {Bezoui}, S. and {Piekarewicz}, J. and {Viens}, F.},
        title = "{Bayesian mixture model approach to quantifying the empirical nuclear saturation point}",
      journal = {\prc},
         year = 2024,
        month = oct,
       volume = {110},
       number = {4},
          eid = {044320},
        pages = {044320},
          doi = {10.1103/PhysRevC.110.044320},
archivePrefix = {arXiv},
       eprint = {2405.02748},
 primaryClass = {nucl-th},
       adsurl = {https://ui.adsabs.harvard.edu/abs/2024PhRvC.110d4320D}
}

@ARTICLE{2019ApJS..241...27A,
       author = {{Ashton}, Gregory and {H{\"u}bner}, Moritz and {Lasky}, Paul D. and {Talbot}, Colm and {Ackley}, Kendall and {Biscoveanu}, Sylvia and {Chu}, Qi and {Divakarla}, Atul and {Easter}, Paul J. and {Goncharov}, Boris and {Hernandez Vivanco}, Francisco and {Harms}, Jan and {Lower}, Marcus E. and {Meadors}, Grant D. and {Melchor}, Denyz and {Payne}, Ethan and {Pitkin}, Matthew D. and {Powell}, Jade and {Sarin}, Nikhil and {Smith}, Rory J.~E. and {Thrane}, Eric},
        title = "{BILBY: A User-friendly Bayesian Inference Library for Gravitational-wave Astronomy}",
      journal = {\apjs},
         year = 2019,
        month = apr,
       volume = {241},
       number = {2},
          eid = {27},
        pages = {27},
          doi = {10.3847/1538-4365/ab06fc},
archivePrefix = {arXiv},
       eprint = {1811.02042},
 primaryClass = {astro-ph.IM},
       adsurl = {https://ui.adsabs.harvard.edu/abs/2019ApJS..241...27A}
}

@ARTICLE{2020MNRAS.493.3132S,
       author = {{Speagle}, Joshua S.},
        title = "{DYNESTY: a dynamic nested sampling package for estimating Bayesian posteriors and evidences}",
      journal = {\mnras},
         year = 2020,
        month = apr,
       volume = {493},
       number = {3},
        pages = {3132-3158},
          doi = {10.1093/mnras/staa278},
archivePrefix = {arXiv},
       eprint = {1904.02180},
 primaryClass = {astro-ph.IM},
       adsurl = {https://ui.adsabs.harvard.edu/abs/2020MNRAS.493.3132S}
}

@ARTICLE{2025PhRvD.112h3009T,
       author = {{Tang}, Shao-Peng and {Huang}, Yong-Jia and {Fan}, Yi-Zhong},
        title = "{Phase transition and nuclear symmetry energy from neutron star observations: Constraints in light of PSR J0614-3329}",
      journal = {\prd},
         year = 2025,
        month = oct,
       volume = {112},
       number = {8},
          eid = {083009},
        pages = {083009},
          doi = {10.1103/bmsk-8n85},
archivePrefix = {arXiv},
       eprint = {2507.10025},
 primaryClass = {astro-ph.HE},
       adsurl = {https://ui.adsabs.harvard.edu/abs/2025PhRvD.112h3009T}
}

@ARTICLE{2025ApJ...995...60M,
       author = {{Mauviard}, Lucien and {Guillot}, Sebastien and {Salmi}, Tuomo and {Choudhury}, Devarshi and {Dorsman}, Bas and {Gonz{\'a}lez-Caniulef}, Denis and {Hoogkamer}, Mariska and {Huppenkothen}, Daniela and {Kazantsev}, Christine and {Kini}, Yves and {Olive}, Jean-Francois and {Stammler}, Pierre and {Watts}, Anna L. and {Mendes}, Melissa and {Rutherford}, Nathan and {Schwenk}, Achim and {Svensson}, Isak and {Bogdanov}, Slavko and {Kerr}, Matthew and {Ray}, Paul S. and {Guillemot}, Lucas and {Cognard}, Isma{\"e}l and {Theureau}, Gilles},
        title = "{A NICER View of the 1.4 M$_{{\ensuremath{\odot}}}$ Edge-on Pulsar PSR J0614-3329}",
      journal = {\apj},
         year = 2025,
        month = dec,
       volume = {995},
       number = {1},
          eid = {60},
        pages = {60},
          doi = {10.3847/1538-4357/ae145d},
archivePrefix = {arXiv},
       eprint = {2506.14883},
 primaryClass = {astro-ph.HE},
       adsurl = {https://ui.adsabs.harvard.edu/abs/2025ApJ...995...60M}
}

@ARTICLE{2025PhRvD.112f3003L,
       author = {{Legred}, Isaac and {Brodie}, Liam and {Haber}, Alexander and {Essick}, Reed and {Chatziioannou}, Katerina},
        title = "{Nonparametric extensions of nuclear equations of state: Probing the breakdown scale of relativistic mean-field theory}",
      journal = {\prd},
         year = 2025,
        month = sep,
       volume = {112},
       number = {6},
          eid = {063003},
        pages = {063003},
          doi = {10.1103/9kh9-xfpd},
archivePrefix = {arXiv},
       eprint = {2505.07677},
 primaryClass = {nucl-th},
       adsurl = {https://ui.adsabs.harvard.edu/abs/2025PhRvD.112f3003L}
}

@ARTICLE{2001A&A...380..151D,
       author = {{Douchin}, F. and {Haensel}, P.},
        title = "{A unified equation of state of dense matter and neutron star structure}",
      journal = {\aap},
         year = 2001,
        month = dec,
       volume = {380},
        pages = {151-167},
          doi = {10.1051/0004-6361:20011402},
archivePrefix = {arXiv},
       eprint = {astro-ph/0111092},
 primaryClass = {astro-ph},
       adsurl = {https://ui.adsabs.harvard.edu/abs/2001A&A...380..151D}
}

@ARTICLE{2024PhRvC.109b5807P,
       author = {{Pang}, Peter T.~H. and {Sivertsen}, Lars and {Somasundaram}, Rahul and {Dietrich}, Tim and {Sen}, Srimoyee and {Tews}, Ingo and {Coughlin}, Michael W. and {Van Den Broeck}, Chris},
        title = "{Probing quarkyonic matter in neutron stars with the Bayesian nuclear-physics multimessenger astrophysics framework}",
      journal = {\prc},
         year = 2024,
        month = feb,
       volume = {109},
       number = {2},
          eid = {025807},
        pages = {025807},
          doi = {10.1103/PhysRevC.109.025807},
archivePrefix = {arXiv},
       eprint = {2308.15067},
 primaryClass = {nucl-th},
       adsurl = {https://ui.adsabs.harvard.edu/abs/2024PhRvC.109b5807P}
}

@ARTICLE{1992Natur.357..472U,
       author = {{Usov}, V.~V.},
        title = "{Millisecond pulsars with extremely strong magnetic fields as a cosmological source of {\ensuremath{\gamma}}-ray bursts}",
      journal = {\nat},
         year = 1992,
        month = jun,
       volume = {357},
       number = {6378},
        pages = {472-474},
          doi = {10.1038/357472a0},
       adsurl = {https://ui.adsabs.harvard.edu/abs/1992Natur.357..472U}
}

@ARTICLE{1992ApJ...392L...9D,
       author = {{Duncan}, Robert C. and {Thompson}, Christopher},
        title = "{Formation of Very Strongly Magnetized Neutron Stars: Implications for Gamma-Ray Bursts}",
      journal = {\apjl},
         year = 1992,
        month = jun,
       volume = {392},
        pages = {L9},
          doi = {10.1086/186413},
       adsurl = {https://ui.adsabs.harvard.edu/abs/1992ApJ...392L...9D}
}

@ARTICLE{1998ApJ...505L.113K,
       author = {{Klu{\'z}niak}, W. and {Ruderman}, M.},
        title = "{The Central Engine of Gamma-Ray Bursters}",
      journal = {\apjl},
         year = 1998,
        month = oct,
       volume = {505},
       number = {2},
        pages = {L113-L117},
          doi = {10.1086/311622},
archivePrefix = {arXiv},
       eprint = {astro-ph/9712320},
 primaryClass = {astro-ph},
       adsurl = {https://ui.adsabs.harvard.edu/abs/1998ApJ...505L.113K}
}

@ARTICLE{1998A&A...333L..87D,
       author = {{Dai}, Z.~G. and {Lu}, T.},
        title = "{Gamma-ray burst afterglows and evolution of postburst fireballs with energy injection from strongly magnetic millisecond pulsars}",
      journal = {\aap},
         year = 1998,
        month = may,
       volume = {333},
        pages = {L87-L90},
          doi = {10.48550/arXiv.astro-ph/9810402},
archivePrefix = {arXiv},
       eprint = {astro-ph/9810402},
 primaryClass = {astro-ph},
       adsurl = {https://ui.adsabs.harvard.edu/abs/1998A&A...333L..87D}
}

@ARTICLE{2006ChJAA...6..513G,
       author = {{Gao}, Wei-Hong and {Fan}, Yi-Zhong},
        title = "{Short-living Supermassive Magnetar Model for the Early X-ray Flares Following Short GRBs}",
      journal = {\cjaa},
         year = 2006,
        month = oct,
       volume = {6},
       number = {5},
        pages = {513-516},
          doi = {10.1088/1009-9271/6/5/01},
archivePrefix = {arXiv},
       eprint = {astro-ph/0512646},
 primaryClass = {astro-ph},
       adsurl = {https://ui.adsabs.harvard.edu/abs/2006ChJAA...6..513G}
}

@ARTICLE{2010ApJ...717..245K,
       author = {{Kasen}, Daniel and {Bildsten}, Lars},
        title = "{Supernova Light Curves Powered by Young Magnetars}",
      journal = {\apj},
         year = 2010,
        month = jul,
       volume = {717},
       number = {1},
        pages = {245-249},
          doi = {10.1088/0004-637X/717/1/245},
archivePrefix = {arXiv},
       eprint = {0911.0680},
 primaryClass = {astro-ph.HE},
       adsurl = {https://ui.adsabs.harvard.edu/abs/2010ApJ...717..245K}
}

@ARTICLE{2010ApJ...719L.204W,
       author = {{Woosley}, S.~E.},
        title = "{Bright Supernovae from Magnetar Birth}",
      journal = {\apjl},
         year = 2010,
        month = aug,
       volume = {719},
       number = {2},
        pages = {L204-L207},
          doi = {10.1088/2041-8205/719/2/L204},
archivePrefix = {arXiv},
       eprint = {0911.0698},
 primaryClass = {astro-ph.HE},
       adsurl = {https://ui.adsabs.harvard.edu/abs/2010ApJ...719L.204W}
}

@ARTICLE{2011PhRvL.107e1102S,
       author = {{Sekiguchi}, Yuichiro and {Kiuchi}, Kenta and {Kyutoku}, Koutarou and {Shibata}, Masaru},
        title = "{Gravitational Waves and Neutrino Emission from the Merger of Binary Neutron Stars}",
      journal = {\prl},
         year = 2011,
        month = jul,
       volume = {107},
       number = {5},
          eid = {051102},
        pages = {051102},
          doi = {10.1103/PhysRevLett.107.051102},
archivePrefix = {arXiv},
       eprint = {1105.2125},
 primaryClass = {gr-qc},
       adsurl = {https://ui.adsabs.harvard.edu/abs/2011PhRvL.107e1102S}
}

@ARTICLE{2017RPPh...80i6901B,
       author = {{Baiotti}, Luca and {Rezzolla}, Luciano},
        title = "{Binary neutron star mergers: a review of Einstein{\textquoteright}s richest laboratory}",
      journal = {Reports on Progress in Physics},
         year = 2017,
        month = sep,
       volume = {80},
       number = {9},
          eid = {096901},
        pages = {096901},
          doi = {10.1088/1361-6633/aa67bb},
archivePrefix = {arXiv},
       eprint = {1607.03540},
 primaryClass = {gr-qc},
       adsurl = {https://ui.adsabs.harvard.edu/abs/2017RPPh...80i6901B}
}

@ARTICLE{2022ApJ...934L..17R,
       author = {{Romani}, Roger W. and {Kandel}, D. and {Filippenko}, Alexei V. and {Brink}, Thomas G. and {Zheng}, WeiKang},
        title = "{PSR J0952-0607: The Fastest and Heaviest Known Galactic Neutron Star}",
      journal = {\apjl},
         year = 2022,
        month = aug,
       volume = {934},
       number = {2},
          eid = {L17},
        pages = {L17},
          doi = {10.3847/2041-8213/ac8007},
archivePrefix = {arXiv},
       eprint = {2207.05124},
 primaryClass = {astro-ph.HE},
       adsurl = {https://ui.adsabs.harvard.edu/abs/2022ApJ...934L..17R}
}

@ARTICLE{2020ApJ...896L..44A,
       author = {{Abbott}, R. and {Abbott}, T.~D. and {Abraham}, S. and others},
        title = "{GW190814: Gravitational Waves from the Coalescence of a 23 Solar Mass Black Hole with a 2.6 Solar Mass Compact Object}",
      journal = {\apjl},
         year = 2020,
        month = jun,
       volume = {896},
       number = {2},
          eid = {L44},
        pages = {L44},
          doi = {10.3847/2041-8213/ab960f},
archivePrefix = {arXiv},
       eprint = {2006.12611},
 primaryClass = {astro-ph.HE},
       adsurl = {https://ui.adsabs.harvard.edu/abs/2020ApJ...896L..44A}
}

@ARTICLE{2023PhRvX..13d1039A,
       author = {{Abbott}, R. and {Abbott}, T.~D. and {Acernese}, F. and others},
        title = "{GWTC-3: Compact Binary Coalescences Observed by LIGO and Virgo during the Second Part of the Third Observing Run}",
      journal = {Physical Review X},
         year = 2023,
        month = oct,
       volume = {13},
       number = {4},
          eid = {041039},
        pages = {041039},
          doi = {10.1103/PhysRevX.13.041039},
archivePrefix = {arXiv},
       eprint = {2111.03606},
 primaryClass = {gr-qc},
       adsurl = {https://ui.adsabs.harvard.edu/abs/2023PhRvX..13d1039A}
}

@ARTICLE{2001ApJ...552L..35Z,
       author = {{Zhang}, Bing and {M{\'e}sz{\'a}ros}, Peter},
        title = "{Gamma-Ray Burst Afterglow with Continuous Energy Injection: Signature of a Highly Magnetized Millisecond Pulsar}",
      journal = {\apjl},
         year = 2001,
        month = may,
       volume = {552},
       number = {1},
        pages = {L35-L38},
          doi = {10.1086/320255},
archivePrefix = {arXiv},
       eprint = {astro-ph/0011133},
 primaryClass = {astro-ph},
       adsurl = {https://ui.adsabs.harvard.edu/abs/2001ApJ...552L..35Z}
}

@ARTICLE{2015Natur.523..189G,
       author = {{Greiner}, Jochen and {Mazzali}, Paolo A. and {Kann}, D. Alexander and others},
        title = "{A very luminous magnetar-powered supernova associated with an ultra-long {\ensuremath{\gamma}}-ray burst}",
      journal = {\nat},
         year = 2015,
        month = jul,
       volume = {523},
       number = {7559},
        pages = {189-192},
          doi = {10.1038/nature14579},
archivePrefix = {arXiv},
       eprint = {1509.03279},
 primaryClass = {astro-ph.HE},
       adsurl = {https://ui.adsabs.harvard.edu/abs/2015Natur.523..189G}
}

@ARTICLE{2022PhRvL.128t2701K,
       author = {{Komoltsev}, Oleg and {Kurkela}, Aleksi},
        title = "{How Perturbative QCD Constrains the Equation of State at Neutron-Star Densities}",
      journal = {\prl},
         year = 2022,
        month = may,
       volume = {128},
       number = {20},
          eid = {202701},
        pages = {202701},
          doi = {10.1103/PhysRevLett.128.202701},
archivePrefix = {arXiv},
       eprint = {2111.05350},
 primaryClass = {nucl-th},
       adsurl = {https://ui.adsabs.harvard.edu/abs/2022PhRvL.128t2701K}
}

@ARTICLE{2021PhRvL.127p2003G,
       author = {{Gorda}, Tyler and {Kurkela}, Aleksi and {Paatelainen}, Risto and {S{\"a}ppi}, Saga and {Vuorinen}, Aleksi},
        title = "{Soft Interactions in Cold Quark Matter}",
      journal = {\prl},
         year = 2021,
        month = oct,
       volume = {127},
       number = {16},
          eid = {162003},
        pages = {162003},
          doi = {10.1103/PhysRevLett.127.162003},
archivePrefix = {arXiv},
       eprint = {2103.05658},
 primaryClass = {hep-ph},
       adsurl = {https://ui.adsabs.harvard.edu/abs/2021PhRvL.127p2003G}
}

@ARTICLE{2018MNRAS.478.1377A,
       author = {{Alsing}, Justin and {Silva}, Hector O. and {Berti}, Emanuele},
        title = "{Evidence for a maximum mass cut-off in the neutron star mass distribution and constraints on the equation of state}",
      journal = {\mnras},
         year = 2018,
        month = jul,
       volume = {478},
       number = {1},
        pages = {1377-1391},
          doi = {10.1093/mnras/sty1065},
archivePrefix = {arXiv},
       eprint = {1709.07889},
 primaryClass = {astro-ph.HE},
       adsurl = {https://ui.adsabs.harvard.edu/abs/2018MNRAS.478.1377A}
}

@ARTICLE{2020PhRvD.102f3006S,
       author = {{Shao}, Dong-Sheng and {Tang}, Shao-Peng and {Jiang}, Jin-Liang and {Fan}, Yi-Zhong},
        title = "{Maximum mass cutoff in the neutron star mass distribution and the prospect of forming supramassive objects in the double neutron star mergers}",
      journal = {\prd},
         year = 2020,
        month = sep,
       volume = {102},
       number = {6},
          eid = {063006},
        pages = {063006},
          doi = {10.1103/PhysRevD.102.063006},
archivePrefix = {arXiv},
       eprint = {2009.04275},
 primaryClass = {astro-ph.HE},
       adsurl = {https://ui.adsabs.harvard.edu/abs/2020PhRvD.102f3006S}
}

@ARTICLE{2024ApJ...961...62V,
       author = {{Vinciguerra}, Serena and {Salmi}, Tuomo and {Watts}, Anna L. and {Choudhury}, Devarshi and {Riley}, Thomas E. and {Ray}, Paul S. and {Bogdanov}, Slavko and {Kini}, Yves and {Guillot}, Sebastien and {Chakrabarty}, Deepto and {Ho}, Wynn C.~G. and {Huppenkothen}, Daniela and {Morsink}, Sharon M. and {Wadiasingh}, Zorawar and {Wolff}, Michael T.},
        title = "{An Updated Mass-Radius Analysis of the 2017-2018 NICER Data Set of PSR J0030+0451}",
      journal = {\apj},
         year = 2024,
        month = jan,
       volume = {961},
       number = {1},
          eid = {62},
        pages = {62},
          doi = {10.3847/1538-4357/acfb83},
archivePrefix = {arXiv},
       eprint = {2308.09469},
 primaryClass = {astro-ph.HE},
       adsurl = {https://ui.adsabs.harvard.edu/abs/2024ApJ...961...62V}
}

@ARTICLE{2021Univ....7..182L,
       author = {{Li}, Bao-An and {Cai}, Bao-Jun and {Xie}, Wen-Jie and {Zhang}, Nai-Bo},
        title = "{Progress in Constraining Nuclear Symmetry Energy Using Neutron Star Observables Since GW170817}",
      journal = {Universe},
         year = 2021,
        month = jun,
       volume = {7},
       number = {6},
          eid = {182},
        pages = {182},
          doi = {10.3390/universe7060182},
archivePrefix = {arXiv},
       eprint = {2105.04629},
 primaryClass = {nucl-th},
       adsurl = {https://ui.adsabs.harvard.edu/abs/2021Univ....7..182L}
}

@ARTICLE{2020MNRAS.499L..82M,
       author = {{Most}, Elias R. and {Papenfort}, L. Jens and {Weih}, Lukas R. and {Rezzolla}, Luciano},
        title = "{A lower bound on the maximum mass if the secondary in GW190814 was once a rapidly spinning neutron star}",
      journal = {\mnras},
         year = 2020,
        month = dec,
       volume = {499},
       number = {1},
        pages = {L82-L86},
          doi = {10.1093/mnrasl/slaa168},
archivePrefix = {arXiv},
       eprint = {2006.14601},
 primaryClass = {astro-ph.HE},
       adsurl = {https://ui.adsabs.harvard.edu/abs/2020MNRAS.499L..82M}
}

@ARTICLE{2021ApJ...921..180H,
       author = {{Hsu}, Brian and {Hosseinzadeh}, Griffin and {Berger}, Edo},
        title = "{Magnetar Models of Superluminous Supernovae from the Dark Energy Survey: Exploring Redshift Evolution}",
      journal = {\apj},
         year = 2021,
        month = nov,
       volume = {921},
       number = {2},
          eid = {180},
        pages = {180},
          doi = {10.3847/1538-4357/ac1aca},
archivePrefix = {arXiv},
       eprint = {2104.09639},
 primaryClass = {astro-ph.HE},
       adsurl = {https://ui.adsabs.harvard.edu/abs/2021ApJ...921..180H}
}

@ARTICLE{2024ApJ...971L..20C,
       author = {{Choudhury}, Devarshi and {Salmi}, Tuomo and {Vinciguerra}, Serena and {Riley}, Thomas E. and {Kini}, Yves and {Watts}, Anna L. and {Dorsman}, Bas and {Bogdanov}, Slavko and {Guillot}, Sebastien and {Ray}, Paul S. and {Reardon}, Daniel J. and {Remillard}, Ronald A. and {Bilous}, Anna V. and {Huppenkothen}, Daniela and {Lattimer}, James M. and {Rutherford}, Nathan and {Arzoumanian}, Zaven and {Gendreau}, Keith C. and {Morsink}, Sharon M. and {Ho}, Wynn C.~G.},
        title = "{A NICER View of the Nearest and Brightest Millisecond Pulsar: PSR J0437{\textendash}4715}",
      journal = {\apjl},
         year = 2024,
        month = aug,
       volume = {971},
       number = {1},
          eid = {L20},
        pages = {L20},
          doi = {10.3847/2041-8213/ad5a6f},
archivePrefix = {arXiv},
       eprint = {2407.06789},
 primaryClass = {astro-ph.HE},
       adsurl = {https://ui.adsabs.harvard.edu/abs/2024ApJ...971L..20C}
}

@ARTICLE{2024arXiv241103427S,
       author = {{Sneppen}, Albert and {Just}, Oliver and {Bauswein}, Andreas and {Damgaard}, Rasmus and {Watson}, Darach and {Shingles}, Luke J. and {Collins}, Christine E. and {Sim}, Stuart A. and {Xiong}, Zewei and {Martinez-Pinedo}, Gabriel and {Soultanis}, Theodoros and {Vijayan}, Vimal},
        title = "{Helium as an Indicator of the Neutron-Star Merger Remnant Lifetime and its Potential for Equation of State Constraints}",
      journal = {arXiv e-prints},
         year = 2024,
        month = nov,
          eid = {arXiv:2411.03427},
        pages = {arXiv:2411.03427},
          doi = {10.48550/arXiv.2411.03427},
archivePrefix = {arXiv},
       eprint = {2411.03427},
 primaryClass = {astro-ph.HE},
       adsurl = {https://ui.adsabs.harvard.edu/abs/2024arXiv241103427S}
}

@ARTICLE{2018ApJ...861..114Y,
       author = {{Yu}, Yun-Wei and {Liu}, Liang-Duan and {Dai}, Zi-Gao},
        title = "{A Long-lived Remnant Neutron Star after GW170817 Inferred from Its Associated Kilonova}",
      journal = {\apj},
         year = 2018,
        month = jul,
       volume = {861},
       number = {2},
          eid = {114},
        pages = {114},
          doi = {10.3847/1538-4357/aac6e5},
archivePrefix = {arXiv},
       eprint = {1711.01898},
 primaryClass = {astro-ph.HE},
       adsurl = {https://ui.adsabs.harvard.edu/abs/2018ApJ...861..114Y}
}

@ARTICLE{2013PhRvD..88f7304F,
       author = {{Fan}, Yi-Zhong and {Wu}, Xue-Feng and {Wei}, Da-Ming},
        title = "{Signature of gravitational wave radiation in afterglows of short gamma-ray bursts?}",
      journal = {\prd},
         year = 2013,
        month = sep,
       volume = {88},
       number = {6},
          eid = {067304},
        pages = {067304},
          doi = {10.1103/PhysRevD.88.067304},
archivePrefix = {arXiv},
       eprint = {1302.3328},
 primaryClass = {astro-ph.HE},
       adsurl = {https://ui.adsabs.harvard.edu/abs/2013PhRvD..88f7304F}
}

@ARTICLE{2019Natur.568..198X,
       author = {{Xue}, Y.~Q. and {Zheng}, X.~C. and {Li}, Y. and {Brandt}, W.~N. and {Zhang}, B. and {Luo}, B. and {Zhang}, B. -B. and {Bauer}, F.~E. and {Sun}, H. and {Lehmer}, B.~D. and {Wu}, X. -F. and {Yang}, G. and {Kong}, X. and {Li}, J.~Y. and {Sun}, M.~Y. and {Wang}, J. -X. and {Vito}, F.},
        title = "{A magnetar-powered X-ray transient as the aftermath of a binary neutron-star merger}",
      journal = {\nat},
         year = 2019,
        month = apr,
       volume = {568},
       number = {7751},
        pages = {198-201},
          doi = {10.1038/s41586-019-1079-5},
archivePrefix = {arXiv},
       eprint = {1904.05368},
 primaryClass = {astro-ph.HE},
       adsurl = {https://ui.adsabs.harvard.edu/abs/2019Natur.568..198X}
}

@ARTICLE{2017ApJ...835..181L,
       author = {{L{\"u}}, Hou-Jun and {Zhang}, Hai-Ming and {Zhong}, Shu-Qing and {Hou}, Shu-Jin and {Sun}, Hui and {Rice}, Jared and {Liang}, En-Wei},
        title = "{Magnetar Central Engine and Possible Gravitational Wave Emission of Nearby Short GRB 160821B}",
      journal = {\apj},
         year = 2017,
        month = feb,
       volume = {835},
       number = {2},
          eid = {181},
        pages = {181},
          doi = {10.3847/1538-4357/835/2/181},
archivePrefix = {arXiv},
       eprint = {1612.05691},
 primaryClass = {astro-ph.HE},
       adsurl = {https://ui.adsabs.harvard.edu/abs/2017ApJ...835..181L}
}

@ARTICLE{2013MNRAS.430.1061R,
       author = {{Rowlinson}, A. and {O'Brien}, P.~T. and {Metzger}, B.~D. and {Tanvir}, N.~R. and {Levan}, A.~J.},
        title = "{Signatures of magnetar central engines in short GRB light curves}",
      journal = {\mnras},
         year = 2013,
        month = apr,
       volume = {430},
       number = {2},
        pages = {1061-1087},
          doi = {10.1093/mnras/sts683},
archivePrefix = {arXiv},
       eprint = {1301.0629},
 primaryClass = {astro-ph.HE},
       adsurl = {https://ui.adsabs.harvard.edu/abs/2013MNRAS.430.1061R}
}

@ARTICLE{2006Sci...312..719P,
       author = {{Price}, D.~J. and {Rosswog}, S.},
        title = "{Producing Ultrastrong Magnetic Fields in Neutron Star Mergers}",
      journal = {Science},
         year = 2006,
        month = may,
       volume = {312},
       number = {5774},
        pages = {719-722},
          doi = {10.1126/science.1125201},
archivePrefix = {arXiv},
       eprint = {astro-ph/0603845},
 primaryClass = {astro-ph},
       adsurl = {https://ui.adsabs.harvard.edu/abs/2006Sci...312..719P}
}

@ARTICLE{2019PhRvX...9a1001A,
       author = {{Abbott}, B.~P. and {Abbott}, R. and {Abbott}, T.~D. and others},
        title = "{Properties of the Binary Neutron Star Merger GW170817}",
      journal = {Physical Review X},
         year = 2019,
        month = jan,
       volume = {9},
       number = {1},
          eid = {011001},
        pages = {011001},
          doi = {10.1103/PhysRevX.9.011001},
archivePrefix = {arXiv},
       eprint = {1805.11579},
 primaryClass = {gr-qc},
       adsurl = {https://ui.adsabs.harvard.edu/abs/2019PhRvX...9a1001A}
}

@ARTICLE{2024PhRvD.109h3037T,
       author = {{Tang}, Shao-Peng and {Han}, Ming-Zhe and {Huang}, Yong-Jia and {Fan}, Yi-Zhong and {Wei}, Da-Ming},
        title = "{Mass and radius of the most massive neutron star: The probe of the equation of state and perturbative QCD}",
      journal = {\prd},
         year = 2024,
        month = apr,
       volume = {109},
       number = {8},
          eid = {083037},
        pages = {083037},
          doi = {10.1103/PhysRevD.109.083037},
archivePrefix = {arXiv},
       eprint = {2311.13805},
 primaryClass = {astro-ph.HE},
       adsurl = {https://ui.adsabs.harvard.edu/abs/2024PhRvD.109h3037T}
}

@ARTICLE{2024ApJ...974..244T,
       author = {{Tang}, Shao-Peng and {Huang}, Yong-Jia and {Han}, Ming-Zhe and {Fan}, Yi-Zhong},
        title = "{Upper Limit of Sound Speed in Nuclear Matter: A Harmonious Interplay of Transport Calculation and Perturbative Quantum Chromodynamic Constraint}",
      journal = {\apj},
         year = 2024,
        month = oct,
       volume = {974},
       number = {2},
          eid = {244},
        pages = {244},
          doi = {10.3847/1538-4357/ad7503},
archivePrefix = {arXiv},
       eprint = {2404.09563},
 primaryClass = {astro-ph.HE},
       adsurl = {https://ui.adsabs.harvard.edu/abs/2024ApJ...974..244T}
}

@ARTICLE{2024Sci...383..275B,
       author = {{Barr}, Ewan D. and {Dutta}, Arunima and {Freire}, Paulo C.~C. and {Cadelano}, Mario and {Gautam}, Tasha and {Kramer}, Michael and {Pallanca}, Cristina and {Ransom}, Scott M. and {Ridolfi}, Alessandro and {Stappers}, Benjamin W. and {Tauris}, Thomas M. and {Venkatraman Krishnan}, Vivek and {Wex}, Norbert and {Bailes}, Matthew and {Behrend}, Jan and {Buchner}, Sarah and {Burgay}, Marta and {Chen}, Weiwei and {Champion}, David J. and {Chen}, C. -H. Rosie and {Corongiu}, Alessandro and {Geyer}, Marisa and {Men}, Y.~P. and {Padmanabh}, Prajwal Voraganti and {Possenti}, Andrea},
        title = "{A pulsar in a binary with a compact object in the mass gap between neutron stars and black holes}",
      journal = {Science},
         year = 2024,
        month = jan,
       volume = {383},
       number = {6680},
        pages = {275-279},
          doi = {10.1126/science.adg3005},
archivePrefix = {arXiv},
       eprint = {2401.09872},
 primaryClass = {astro-ph.HE},
       adsurl = {https://ui.adsabs.harvard.edu/abs/2024Sci...383..275B}
}

@ARTICLE{1995ApJ...444..306S,
       author = {{Stergioulas}, Nikolaos and {Friedman}, John L.},
        title = "{Comparing Models of Rapidly Rotating Relativistic Stars Constructed by Two Numerical Methods}",
      journal = {\apj},
         year = 1995,
        month = may,
       volume = {444},
        pages = {306},
          doi = {10.1086/175605},
archivePrefix = {arXiv},
       eprint = {astro-ph/9411032},
 primaryClass = {astro-ph},
       adsurl = {https://ui.adsabs.harvard.edu/abs/1995ApJ...444..306S}
}
\bibliographystyle{aasjournal}

\end{document}